\documentstyle[11pt,aaspp4,epsfig]{article}
\slugcomment{\em To appear in Ap.J.}
\lefthead{Watanabe et al.}
\righthead{The Hubble Constant}

\newcommand{\kms}{km\,s$^{-1}$}
\newcommand{\kmsMpc}{km\,s$^{-1}$Mpc$^{-1}$}

\newcommand{\TF}{Tully-Fisher}

\newcommand{\PP}{Pisces-Perseus}

\newcommand{\II}{I\hspace{-0.3ex}I}
\newcommand{\III}{I\hspace{-0.3ex}I\hspace{-0.3ex}I}

\begin{document}
\baselineskip 3ex
\title{An Unbiased Estimate of the Global Hubble Constant \\
       in the Region of Pisces-Perseus}
\author{Masaru Watanabe\altaffilmark{1}}
\affil{Department of Astronomy, University of Tokyo, Tokyo 113, Japan}
\altaffiltext{1}{Present address : National Astronomical Observatory of
Japan, Tokyo 181, Japan}
\author{Takashi Ichikawa\altaffilmark{2}}
\affil{Kiso Observatory, University of Tokyo, Nagano 397-01, Japan}
\altaffiltext{2}{Present address : Astronomical Institute, Tohoku
University, Sendai 980-77, Japan}
\and
\author{Sadanori Okamura}
\affil{Department of Astronomy, University of Tokyo, Tokyo 113, Japan}

\begin{abstract}
We obtain an unbiased estimate of the global Hubble constant $H_0$ in
the volume of $cz\le 12000$\,\kms\ in the region of Pisces-Perseus.
The Tully-Fisher (TF) relation is applied to a magnitude-limited sample 
of 441 spirals selected from the Arecibo 21\,cm catalog. We improve the
photometry data of the previous TF study by Ichikawa and
Fukugita (1992) by using our original surface photometry data
and local calibrators which are newly made available in the literature.
The photometry data were
calibrated with CCD observations and we achieve 0.13\,mag for a
photometric internal error. We use a maximum likelihood method for
the TF analysis. Monte-Carlo simulations demonstrate that our
method reproduces a given $H_0$ at the 95\% confidence level. By 
applying the method to our sample
galaxies, we obtain the unbiased global Hubble constant
$H_0=65\pm 2^{\;+20}_{\;-14}$\,\kmsMpc\ ;
the first and
the second terms represent the internal random error and the external
uppermost and lowermost systematic errors, respectively.
We also find a good
agreement for our $H_0$ with those recently obtained via Cepheid
observation in the local Universe and the TF relation and
supernovae applied to a spatial volume comparable with or larger than
ours.
Hubble velocities of the spirals inferred from our $H_0$ show no
significant systematic difference from those given in the Mark \III\
catalog.
The same analysis for $H_0$ is carried out
using $r$-band photometry data of the \PP\ region given by Willick et
al. (1997). We obtain a global $H_0$
which is consistent with that obtained from the $B$-band analysis.
A bulk motion in the Pisces-Perseus region is briefly discussed, based 
on our calibration of $H_0$. The $B$-band intrinsic TF scatter is too 
large to allow any determination of bulk 
motion. However, our $r$-band TF analysis supports the notion of a 
coherent
streaming motion of the Pisces-Perseus ridge with $\sim -200$\,\kms\ 
with respect to the CMB, in agreement with most modern studies.
\end{abstract}


\section{Introduction} \label{sec:INTRODUCTION}

The global value of the Hubble constant $H_0$ is determined by
investigating a Hubble flow of a large number of galaxies distributed in
a volume larger than the scale of inhomogeneous galaxy distribution in
the Universe ($\sim 100\,h^{-1}$\,Mpc, $h\equiv H_0/100$\,\kmsMpc ).
Ichikawa \& Fukugita (1992, hereafter IF92) determined the global $H_0$
by applying the Tully-Fisher (TF) relation (Tully \& Fisher 1977) to the
Arecibo HI 21\,cm sample (\cite{GH85,GH86,GH89}). The sample extends in
the area of 950 square-degrees in the region of Pisces-Perseus
($22^h\le \alpha_{1950}\le 4^h$,
$+22^\circ\le \delta_{1950}\le +33^\circ$) and includes all the field
and cluster galaxies complete down to $m_z=15.5$\,mag originated from
CGCG (Zwicky et al. 1961-68). Because of the large area and the deep
limiting magnitude, the sample IF92 used contains about 400 spirals at
$cz\le 12000$\,\kms . This large sample enabled them to examine an
effect of biases in the TF analysis.

In this paper, we re-determine $H_0$ using the Arecibo HI sample. The
differences from the study of IF92 are as follows. First,
the imaging data were extracted from our photographic plate material via
modern image reduction techniques.
IF92 used magnitudes and axial ratios which were transformed into a
homogeneous system of RC2 (de Vaucouleurs, de Vaucouleurs \& Corwin
1976) from CGCG, UGC (Nilson 1973), and the Arecibo catalog.
Second, we use a new set of local calibrators for TF calibration.
Their distances are all calibrated with Cepheid observations. Third,
our TF analysis uses a technique of maximum likelihood method.
Fourth, our TF analysis is also
applied to $r$-band CCD data of ``w91pp'' in the Mark \III\ catalog
(Willick et al. 1997). The sample was originated from UGC and hence
complete down to $m_z=14.5$\,mag, shallower than ours by 1\,mag.
Nevertheless, the data allow us to carry out a useful comparison with
our result because the $r$-band TF relation is tighter than that of
$B$-band.
Finally, we discuss, based on our result of the global $H_0$, the bulk
motion of galaxies in the region which has been studies by many authors
(Willick 1990, Han and Mould 1992, Courteau et al. 1993, Hudson et al.
1997).

Our adoption of photographic data may deserve a comment. Motivation of
the present study is to determine $H_0$ in the \PP\ region where a bulk
motion of galaxies has been attracted much attention (Willick 1991 ; Han
and Mould 1992 ; Courteau et al. 1993 ; Hudson et al. 1997). The bulk
motion has been usually computed using relative distances of galaxies
with respect to a reference point of the Hubble flow. In this procedure,
the motion is derived independently from $H_0$. In this paper, on the
other hand, we attempt to determine the global $H_0$ and then discuss
the bulk flow using the absolute value of $H_0$.

To achieve the goal of our study, our galaxy sample should fulfill three
conditions. First, it should be large in terms of the spatial volume so
that the global $H_0$ is determined unaffected by the bulk flow. The
volume is required to be larger than the coherent scale of the flow
($cz\sim 6000$\,\kms ). Second, our sample should be also large in terms
of the number of galaxies and should not be sparse so that we can
evaluate the effect of a bias correctly.
Third, it is of course highly desired that photometry data are
homogeneous so as to eliminate systematic errors which creep in a
data set taken from different compilations. Considering
these conditions, we carried out photographic observations and obtained
homogeneous photometry data for the \PP\ region. The galaxy sample is a
magnitude limited one containing all the CGCG/UGC galaxies, and is
distributed up to $cz\sim 12000$\,\kms . The sample is therefore larger
than a \PP\ field sample ``w91pp'' in the Mark \III\ catalog (Willick
et al. 1997). A recent study of $H_0$ by Giovanelli et al. (1997a)
covered the entire sky using 24 clusters. Although the \PP\ region was
included in their research, their sample in the region is limited to
five clusters including about 150 galaxies. Further,
the accuracy of our photographic photometry is as good as 0.13\,mag
(\cite{Watanabe96}). It is worth stressing that careful photographic
photometry do achieve 0.1-0.2\,mag accuracy (e.g., Kodaira et al. 1990
; Yasuda, Okamura and Fukugita 1995). Thus, our photographic
observations offer the photometry data which are most suitable for
our study.

Outline of this paper is as follows. The data used for the present
analysis are described in \S \ref{sec:DATA}. The
maximum-likelihood method of the TF analysis is presented in \S
\ref{sec:A NEW METHOD OF THE ANALYSIS}. Calibration of the TF relation
is given in \S \ref{sec:THE TULLY-FISHER RELATION}. Reliability of the
present method is examined by Monte-Carlo simulations in \S
\ref{sec:MONTE-CARLO SIMULATIONS}. The method is applied to our $B$-band
sample and the unbiased estimate of the global $H_0$ is determined in \S
\ref{sec:RESULT}.
A comparative analysis is carried out using ``w91pp'' data of Mark \III\
compilation in \S \ref{sec:ANALYSIS WITH WILLICK'S r-BAND SAMPLE}. We
discuss the present result in \S \ref{sec:DISCUSSIONS}.

\section{Data} \label{sec:DATA}
\subsection{Observations and Data Reduction} \label{sec:DATA
Observations and Data Reduction}

Observations and data reduction are briefly described here.
Details will be given in a forthcoming paper.

Photographic observations of the region of \PP\
($22^h\le \alpha_{1950}\le 4^h$
and $+22^\circ\le \delta_{1950}\le +33^\circ$) were carried out in
$B$ (Kodak \II a-O emulsion plus Schott GG385 glass filter) using the
105\,cm Schmidt telescope at the Kiso Observatory, the University of
Tokyo. Most of the observations were conducted in 1991 and in 1992 and
we obtained thirty-eight plates. We measured and reduced all
the 1524 CGCG and UGC galaxies in the region. These 1524 galaxies are
referred to as the photometry sample. The galaxies were scanned with a
PDS2020
microdensitometer at the Kiso Observatory. We then reduced the images
in a conventional manner for Schmidt plate data (e.g., Kodaira et al.
1990) using the software packages IRAF\footnote{{\em IRAF}\/ is
distributed by the National Optical Astronomy Observatories, which is
operated by the Association of Universities for Research in Astronomy,
Inc. (AURA) under cooperative agreement with the National Science
Foundation.} and SPIRAL (Ichikawa et al. 1987\,; Hamabe and Ichikawa
1991). CCD observations were also carried out in $B$ and in $V$ for
seventy spirals selected from the photometry sample to calibrate the
photographic data. The CCD data were calibrated with photometric
standard stars (Landolt 1992) of various colors observed at
various zenith distances to correct atmospheric extinction and to
transform our photographic $B$-band system to the Johnson's. For seven
plates where no CCD observation was available, the flux calibration was
made with photoelectric aperture photometry data of Longo and de
Vaucouleurs (1983). The flux calibration of the photographic data using
the CCD and the photoelectric aperture photometry data followed the
conventional method of Kodaira et al. (1990). Aperture magnitudes were
measured with a series of concentric circular apertures throughout the
present study. Total magnitudes $B_T$ were obtained by a sliding fit of
aperture-magnitude curve (hereafter a growth curve) to their templates
given in RC3 (de Vaucouleurs et al. 1991). Isophotal magnitudes $B_{25}$
and major-to-minor axial ratios $R_{25}$ were measured at the isophote
of 25\,mag\,arcsec$^{-2}$ in the surface brightness. Internal errors
were 0.13\,mag for $B_T$ and 0.06\,dex for $\log R_{25}$
(\cite{Watanabe96}).

Spiral galaxies are selected from the photometry sample and used for the
present TF analysis. The procedure of the selection is described in \S
\ref{sec:DATA Sample selection}.

\subsection{Comparisons with Other Data} \label{sec:DATA Comparison with
Other Data}

\begin{figure}
\vspace*{-30mm}
\begin{center}
\epsfig{file=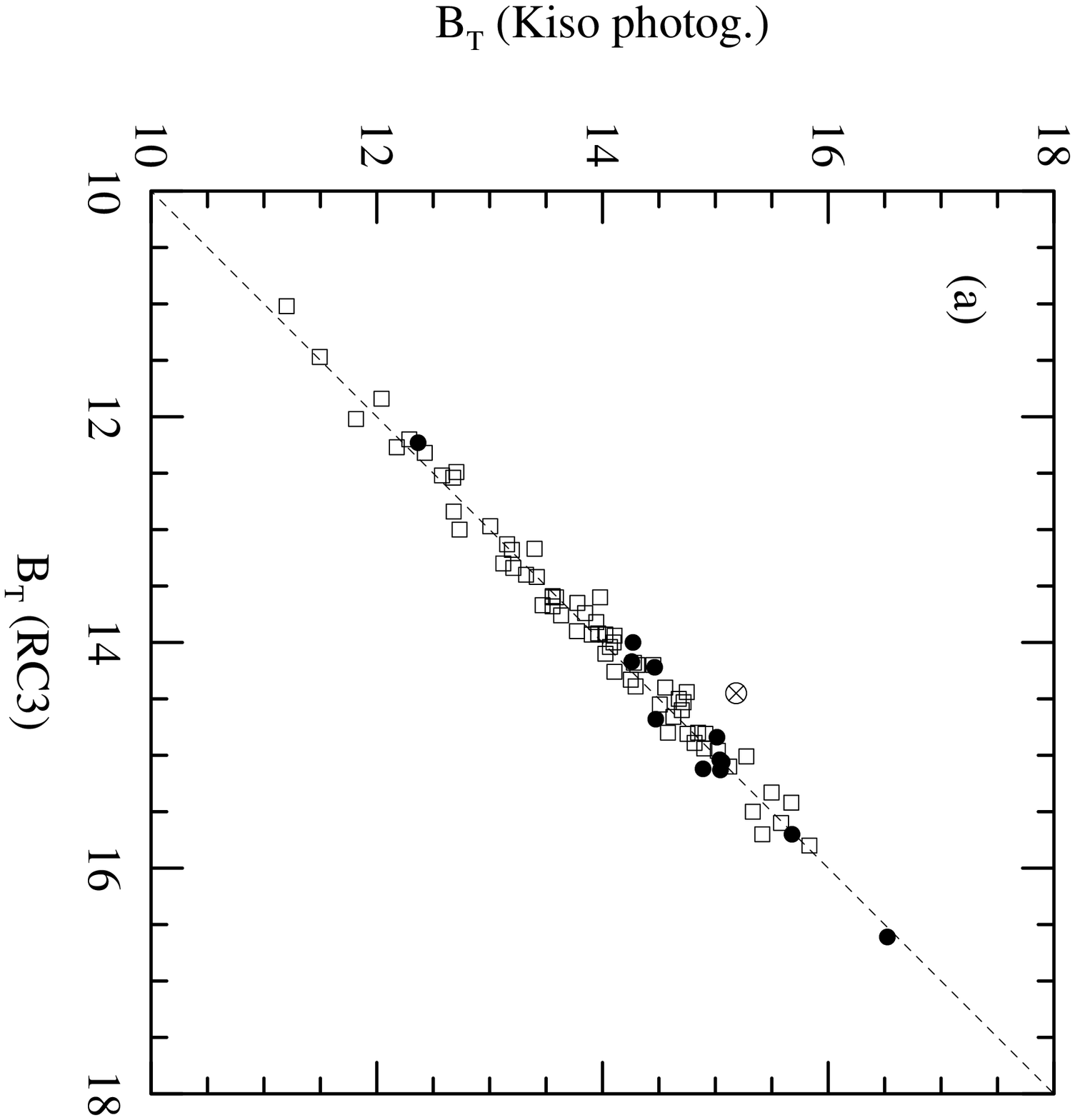, 
        width=350pt, height=300pt, rheight=280pt,
	bbllx=70pt, bburx=770pt, bblly=20pt, bbury=620pt, 
	clip=, angle=0}
\epsfig{file=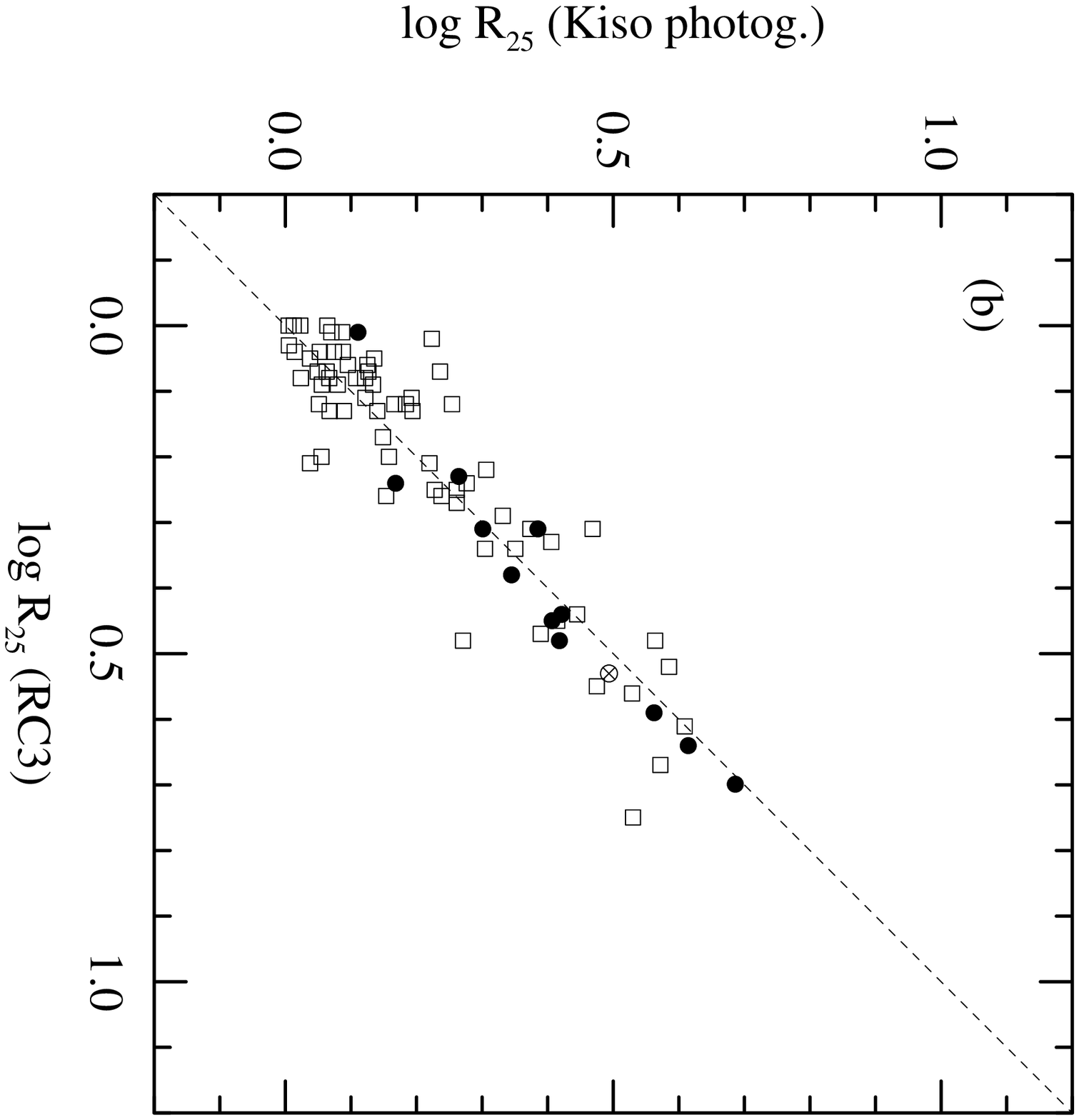, 
        width=350pt, height=300pt,
        bbllx=70pt, bburx=770pt, bblly=20pt, bbury=620pt, 
        clip=, angle=0}
\end{center}\vspace{-5mm}
\caption{Comparisons ($a$) of the total magnitude $B_T$ and ($b$) of
         the logarithmic isophotal major-to-minor axial ratio
	 $\log R_{25}$, between RC3 data and ours. Filled circles
	 represent galaxies for which RC3 data are based on surface
	 photometry. A encircled cross indicates
	 PGC\,5473 (see text). \label{fig:BtRC3_vs_Bt}}
\end{figure}

\begin{figure}
\vspace*{-30mm}
\begin{center}
\epsfig{file=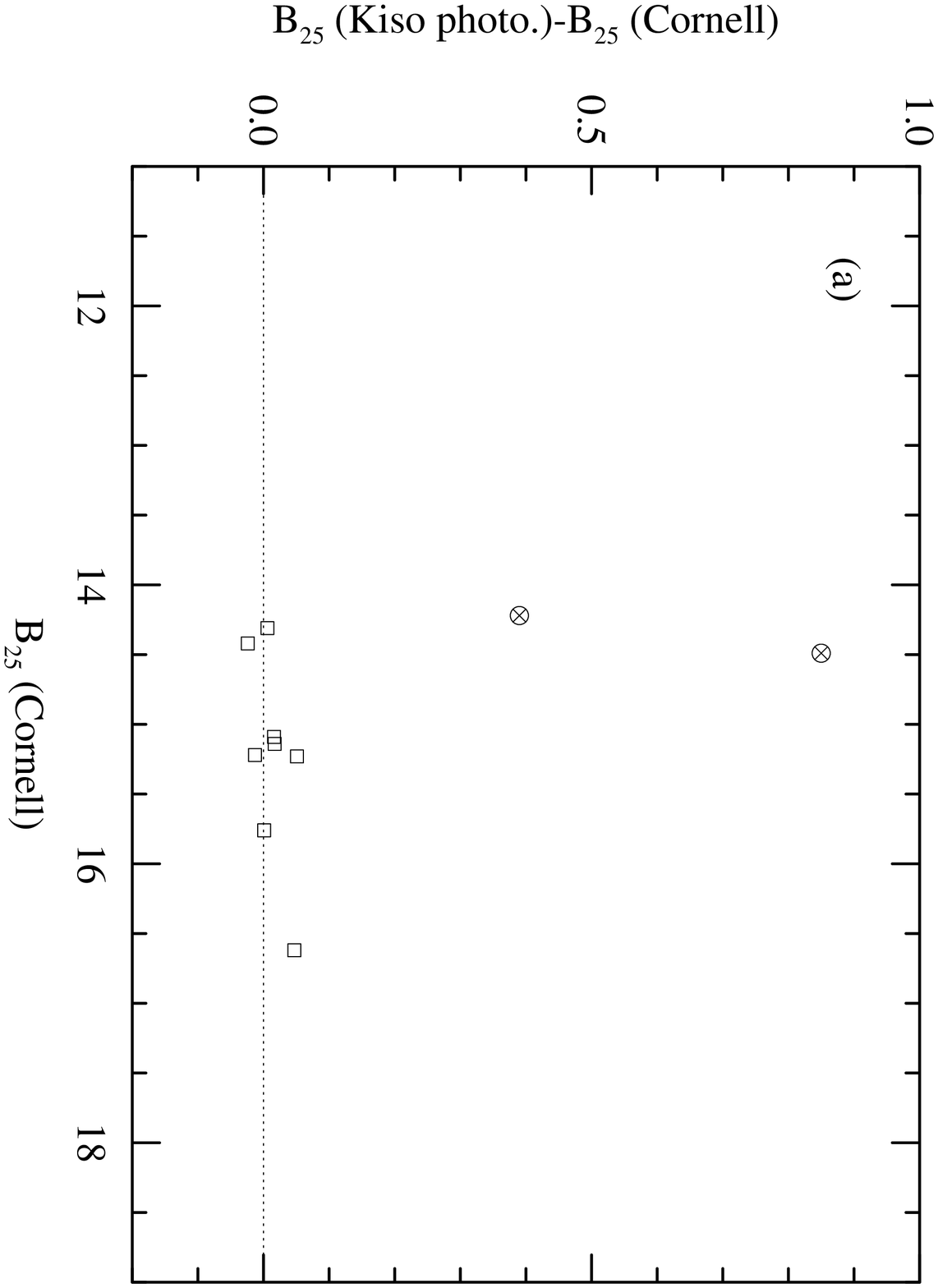,
        width=350pt, height=300pt, rheight=280pt,
	bbllx=70pt, bburx=770pt, bblly=20pt, bbury=620pt, 
	clip=, angle=0}
\epsfig{file=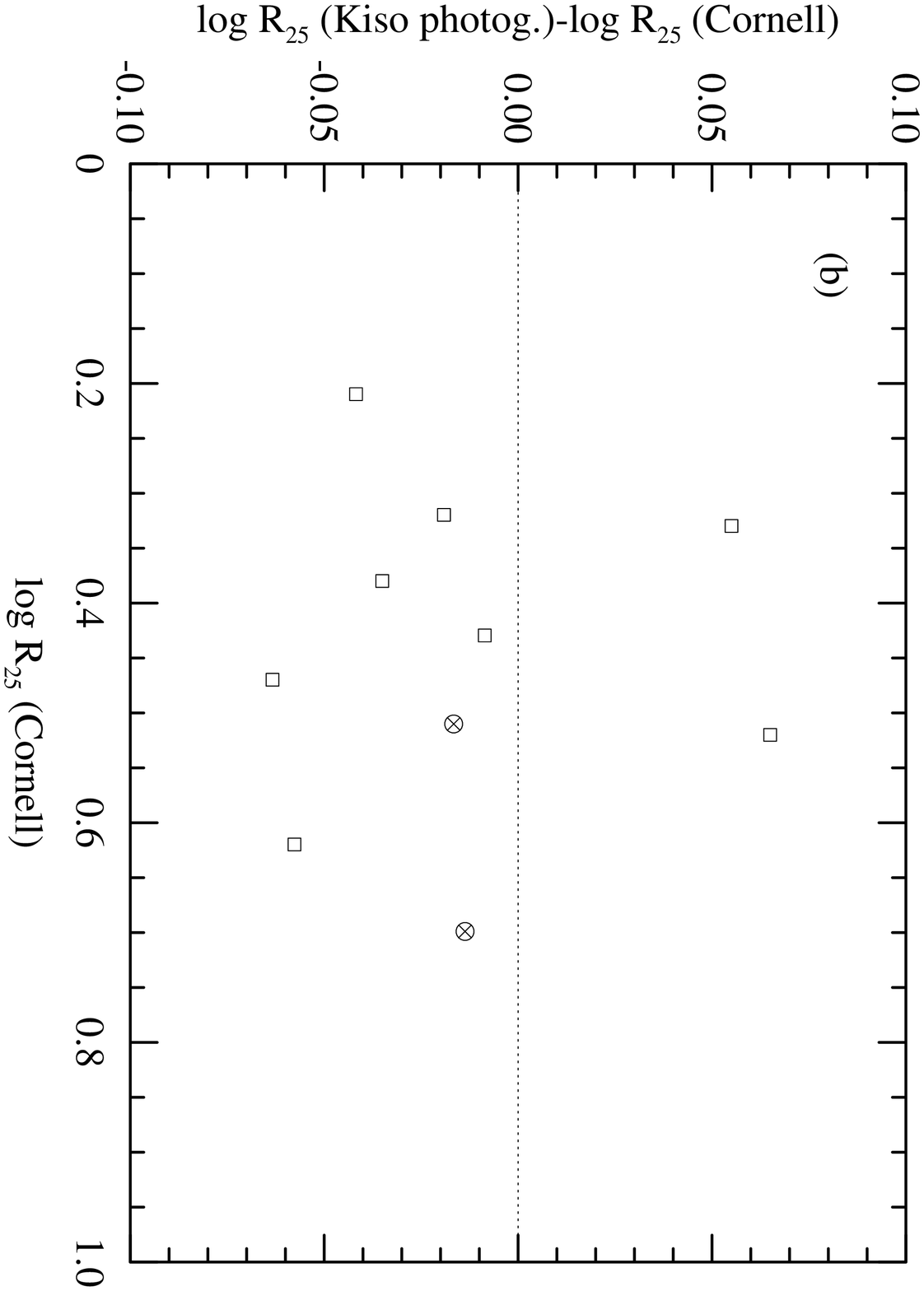,
        width=350pt, height=300pt,
        bbllx=70pt, bburx=770pt, bblly=20pt, bbury=620pt, 
        clip=, angle=0}
\end{center}\vspace{-5mm}
\caption{Comparisons ($a$) of the isophotal magnitude $B_{25}$ and
	 ($b$) of the logarithmic isophotal major-to-minor axial ratio
	 $\log R_{25}$, between the data of Cornell et al (1987) and
	 ours. Encircled crosses
	 indicate PGC\,5473 and PGC\,3133 (see text).
	 \label{fig:B25Cornell_vs_B25}}
\end{figure}

We examine external consistency by comparing our data with those
in RC3 (N=79) and Cornell et al. (1987) (N=8). All overlapping systems
were rejected from the comparisons. We find possible errors in $B_T$ of
RC3 for PGC\,5473 (CGCG\,502-87, UGC\,1045) and in $B_T$ of RC3 and in
$B_{25}$ of Cornell et al. (1987) for PGC\,3133 (Watanabe 1996). These
galaxies are further rejected. The comparisons are presented in
Fig.\ref{fig:BtRC3_vs_Bt} and Fig.\ref{fig:B25Cornell_vs_B25}.
From a regression analysis for RC3 data, we obtain
$B_T\mbox{(Kiso)}=B_T\mbox{(RC3)}+0.02$\,mag ($\sigma=0.16$\,mag) and
$\log R_{25}\mbox{(Kiso)}=\log R_{25}\mbox{(RC3)}+0.00$
($\sigma=0.07$\,dex). If we assume that the half of $\sigma=0.16$\,mag
is ascribed to the error in RC3, our $B_T$ is supposed to be as accurate
as $0.11$\,mag. This is compatible with our internal error of 0.13\,mag
for $B_T$. Rms errors of $B_T$ between RC3 data and ours are 0.17\,mag
(N=46) for galaxies calibrated with our CCD observation while 0.14\,mag
(N=32) for those calibrated with the aperture photometry data by Longo
and de Vaucouleurs (1983). The difference between these two values is
insignificant, and we do not think the calibration by the aperture
photometry data is poorer than that by the CCD data.
A regression analysis for Cornell's data leads to
$B_{25}\mbox{(Kiso)}=B_{25}\mbox{(Cornell)}+0.01$\,mag
($\sigma=0.01$\,mag) and
$\log R_{25}\mbox{(Kiso)}=\log R_{25}\mbox{(Cornell)}-0.01$
($\sigma=0.04$\,dex). These good agreements with their CCD data
demonstrate reliability of our photographic photometry data.

\subsection{HI Data} \label{sec:DATA HI Data}

HI 21\,cm data are taken from the Arecibo HI catalog
(\cite{GH85,GH86,GH89}), where line-profile widths are defined at the
50\% level of the line
mean intensity. The line widths $W_{50}^{obs}$ and heliocentric
recession velocities $V_h$ both have an accuracy of about $10\,$\kms .

In earlier TF calibration, the 21\,cm line-profile width has been
defined with two schemes. One refers to the width $W_{50}^{obs}$ as
used in
the Arecibo catalog, while the other adopted a width $W_{20}^{obs}$
measured at the 20\% level of the line peak intensity. We adopt the
latter definition $W_{20}^{obs}$ because most of the earlier TF
calibrations have been made with this scheme (e.g., Aaronson et al.
1986, Bottinelli et al. 1987, Pierce and Tully 1988, Willick 1990,
Pierce \& Tully 1992, IF92). Adopting this schemes allows us to
compare our TF calibration easily with those of the previous studies.
We convert $W_{50}^{obs}$ to $W_{20}^{obs}$ using an empirical formula
given by Willick (1991) ;

\begin{equation}
\log W_{20}^{obs}-2.5=0.933\,(\log W_{50}^{obs}-2.5)+0.019\,.
\label{eqn:W50toW20}
\end{equation}

\noindent
The relation has a dispersion of $0.02$\,dex. The uncertainty caused by
this dispersion is less than 0.1\,mag which is negligible compared with
the intrinsic dispersion of the TF relation, $\ge 0.3$\,mag.

We determine the global $H_0$ with reference to a rest frame of the
cosmic microwave background (CMB) radiation. The heliocentric velocity
$V_h$ is accordingly transformed to $V_c$ measured in the frame using
CMB dipole anisotropy attributed to the solar motion of $v=365$\,\kms\
toward $(l,b)=(265^\circ, +48^\circ)$ relative to the CMB radiation
field (Smoot et al. 1991). Let $l_g$ and $b_g$ be the galactic
longitude and latitude of a particular galaxy. Then, the transformation
is expressed as
$V_c=V_h+v_x+v_y+v_z$ where
$v_x=v\cos b\cos l\cos b_g\cos l_g$,
$v_y=v\cos b\sin l\cos b_g\sin l_g$, and
$v_z=v\sin b\sin b_g$, respectively.

The Arecibo catalog mainly refers to UGC for galaxy morphological types.
For galaxies for which the type was not available in UGC, authors of
the Arecibo catalog determined the type from visual inspection of
galaxy images on Palomar Schmidt plates or on Palomar Sky Survey
prints. We take their morphological type with a notation $T_{G\!H}$.

\subsection{Correction to the Observational Parameters}
 \label{sec:DATA Correction to the Observables}
\subsubsection{Apparent magnitudes}
 \label{sec:DATA Correction to the Observables Apparent magnitudes}

The observed magnitude $B_T$ is corrected for Galactic and internal
extinction and K-dimming. The Galactic extinction $A_G$ is taken from
a machine-readable extinction map made by David Burstein (hereafter
referred to as BH).
The internal extinction $A_i$ and the K-dimming $K$ are both evaluated
following a prescription given in RC3. Corrected magnitude $B_T^0$ is
then given by $B_T^0=B_T-A_G-A_i-K$\,.

There are various schemes used for evaluation of $A_G$ (e.g., BH, RC2,
RSA [Sandage and Tammann 1981]) and $A_i$ (e.g., RC3, RC2, RSA, Bothun
et al. [1985]). Systematic offset among these different schemes will be
taken into account as a source of an error for $H_0$\,(\S
\ref{sec:RESULT Global Value}).

\subsubsection{Line-profile widths}
 \label{sec:DATA Correction to the Observables Line-profile widths}

The line-profile width $W_{20}^{obs}$ is corrected for a projection and
a cosmological broadening effects with

\begin{equation}
W_{20}=\frac{W_{20}^{obs}}{(1+z)\,\sin i}\,,
\label{eqn:W0 calculation}
\end{equation}

\noindent
where inclination $i$ of a spiral is estimated from $R_{25}$ via a
formula

\begin{equation}
\sin i=\left(\frac{1-R_{25}^{-2}}{1-R_0^{-2}}\right)^{\frac{1}{2}}\,
\label{eqn:R to sini}
\end{equation}

\noindent

Dependence of the intrinsic major-to-minor axial ratio $R_0$ on a galaxy
type was investigated by some authors (e.g.,
\cite{Bottinelli83,Fouque90,Giovanelli97b}). Some authors also favors
a low value $R_0<0.2$ (Giovanelli et al. 1994 ; Giovanelli et al.
1997b ; Willick et al. 1997).
However, we use a conservative value $R_0^{-1}=0.2$ for all spirals
following most of earlier analysis (Pierce and Tully 1988 ; Fukugita et
al. 1991 ; Freudling, Martel and Haynes 1991 ; IF92 ; Han 1992 ; Pierce
and Tully 1992 ; Bernstein et al. 1994 ; Bureau, Mould and
Staveley-Smith 1996). We find that adopting a low value of
$R_0^{-1}=0.13$ or introducing its type dependence results in an
insignificant change in $H_0$ by $\sim 1$\,\kmsMpc\ in our analysis.

Giovanelli et al. (1997b) proposed a different formula for the
correction of the line-profile width $W_{50}$. A similar formula could
be worked out for $W_{20}$. However, we did not pursue this way. The
ratio of their observed width W to the corrected width $W_1$ is almost
constant ; the mean of $W/W_1$ is 1.07 with a scatter of only 0.07.
It is reasonable to assume that the ratio would be nearly constant in
case such a revision is made for $W_{20}$. Further, such a revision
would change $W_{20}$ of local calibrators as well as sample galaxies.
Then, the effect of such a revision is likely to be canceled out.

\subsection{Sample Selection} \label{sec:DATA Sample selection}

We select sample galaxies for our TF analysis according to selection
criteria as follows; 1) Morphological type Sa-Sd, i.e.,
$3\leq T_{G\!H}\leq 8$ and $T_{G\!H}=12$. This criterion is
adopted to exclude S0 and very late-type spirals unsuitable for the TF
analysis. Galaxies classified to $T_{G\!H}=12\,$ are all
regarded as Sb in the present analysis. 2) Inclination
$i\geq 45^{\circ}$. For galaxies with $i<45^{\circ}$, an
uncertainty of the TF magnitude caused by an error of
$\Delta\log R_{25}=0.06$ (Watanabe 1996) combined with a slope of
6.14 for our TF calibration (\S
\ref{sec:THE TULLY-FISHER RELATION Calibration}) exceeds
0.4\,mag, which is no more negligible compared with the intrinsic
dispersion of the TF relation. 3) Recession velocity
$V_c\geq 1000\,$\kms . This criterion is adopted to avoid disturbance
by a local peculiar velocity field with amplitude of several hundreds
\kms . 4) Line-profile width
$2.30\leq\log W_{20}<2.75\,$. As discussed in
\S\ref{sec:THE TULLY-FISHER RELATION Calibration}, the lower and the
upper boundaries are originated from the range of $\log W_{20}$ for the
local calibrators we adopted.

\begin{figure}
\vspace*{-20mm}
\begin{center}
\epsfig{file=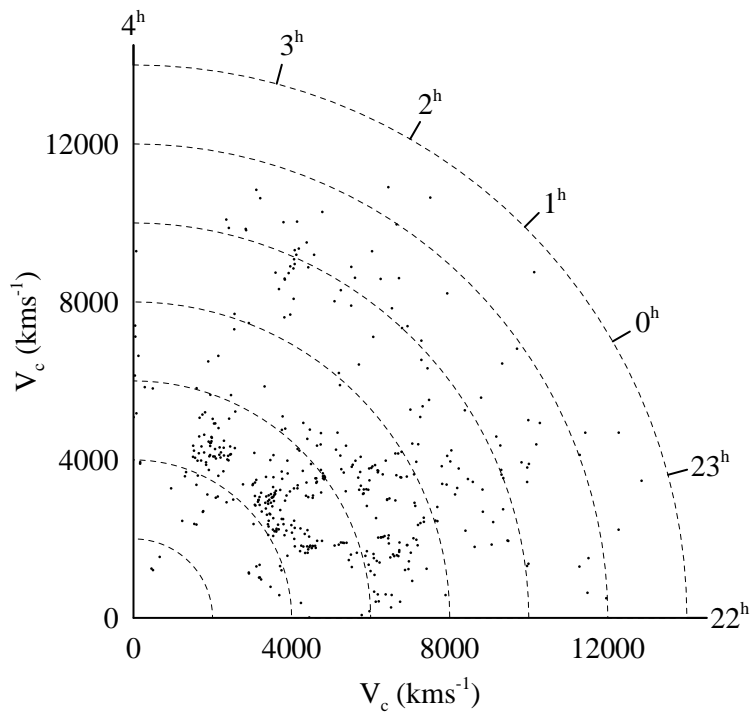,
        width=350pt, height=300pt,
        bbllx=70pt, bburx=770pt, bblly=20pt, bbury=620pt, 
        clip=, angle=0}
\end{center}\vspace{-5mm}
\caption{Velocity distribution of the spiral sample ($N=449$). The
         velocity is expressed in the CMB-rest frame.
	 \label{fig:cz_plot_449}}
\end{figure}

These criteria reduced the photometry sample to 449 galaxies (Table
\ref{tbl:Sample selection for the TF analysis}). These are hereafter
referred to as the spiral sample. The TF analysis is carried out using
441 galaxies among them (\S\ref{sec:RESULT}). The
eight galaxies are discarded from the analysis because of their large
deviation in the TF magnitude; this is discussed in
\S\ref{sec:THE TULLY-FISHER RELATION Consistency between the Local
Calibrators and the TF Sample}. A wedge diagram for the spiral sample is
shown in Fig.\ref{fig:cz_plot_449}, which illustrates that the sample
covers a spatial volume up to $V_c\simeq 12000$\,\kms . Number
distribution as a function of $B_T$ is shown in
Fig.\ref{fig:Bt_vs_N_1524_957_449}. The number ratio of the spiral
sample to the photometry sample is almost constant over
$13\le B_T\le 17$\,mag. The spiral sample is therefore supposed to
be a fair subset of the photometry sample.

\section{A MAXIMUM LIKELIHOOD TECHNIQUE} \label{sec:A NEW METHOD OF THE
ANALYSIS}

If we obtain a TF distance $r_{T\!F}$ using $B_T^0$ and a TF magnitude
$M_{T\!F}$ as $r_{T\!F}=10^{(B_T^0-M_{T\!F}+5)/5}\,\mbox{pc}$, the most
straightforward estimate of
$H_0$ may be a ratio $V_c/r_{T\!F}$. We hereafter call it as the Hubble
ratio. For our spiral sample, however, a sample average of the Hubble
ratios is not a correct estimate of the true global $H_0$. It is subject
to a systematic error because the sample galaxies are biased toward
intrinsically bright ones. This bias is originated from the facts that
the intrinsic dispersion $\sigma_{T\!F}$ of the TF relation is non-zero
and that the spiral sample is a magnitude-limited sample. Consequently,
a systematic error occurs in the sample average of the Hubble ratio. In
this paper, we refer to this bias as a {\em sampling incompleteness}\/
bias. We also take the {\em Malmquist}\/ bias (Malmquist 1922) into
account in the present analysis. The Malmquist bias is caused by a
non-zero intrinsic dispersion of the TF relation combined with a
conical shape of an observed volume for a unit solid angle. A detailed
review for the two biases is found in Strauss and Willick (1995). Note
that an error caused by the Malmquist bias is independent of a
limiting magnitude, while an error caused by the sampling
incompleteness bias depends basically on the limiting magnitude.

To determine a global $H_0$ which is not affected by such systematic
errors, we use a maximum likelihood method in the TF analysis. A similar
technique was used by Han and Mould (1990) for an analysis of a Local
Supercluster velocity field. The likelihood function is computed using
a probability distribution of $W_{20}$ rather than $B_T$ since it is
less affected by the sample incompleteness (e.g., Schechter 1980\,;
Freudling et al. 1995, and references therein).
This is equivalent to the ``inverse method'' discussed by Strauss and
Willick (1995).

\subsection{Formulation}
\label{sec:A NEW METHOD OF THE ANALYSIS Formalism}

Let us consider a probability ${\cal P}^{(i)}$ that the $i$-th spiral
galaxy has a line-profile width $W_{20}^{obs(i)}$ under the condition
that it has a true absolute magnitude $M_B^{(i)}$, an inclination
$i^{(i)}$, and a redshift $z^{(i)}$. Note that $M_B^{(i)}$ is an
unknown parameter. An intrinsic line-profile width $W_{20}^{(i)}$ is
obtained from $W_{20}^{obs(i)}$, $i^{(i)}$, and $z^{(i)}$ with equations
(\ref{eqn:W0 calculation}) and (\ref{eqn:R to sini}). A TF magnitude
$M_{T\!F}^{(i)}$ corresponding to $W_{20}^{(i)}$ is then obtained by
the TF relation,

\begin{equation}
M_{T\!F}^{(i)}=\mbox{A}+\mbox{B}\,(\log W_{20}^{(i)}-2.5)\,.
\label{eqn:general TFrelation}
\end{equation}

\noindent
On the other hand, the Hubble's law gives an estimate for $M_B^{(i)}$,
i.e.,

\begin{equation}
M_B^{(i)}=B_T^{0(i)}-5\log V_c^{(i)}+5\log h-15-\delta\,,
\label{eqn:Mabs_de}
\end{equation}

\noindent
where $V_c^{(i)}$ is given in unit of \kms . The additional term
$\delta$ is a correction for a systematic error caused by the Malmquist
bias. If galaxies are uniformly distributed in space, the systematic
error is $\delta=(3\ln 10/5)\,\sigma^2=1.38\sigma^2\,$, where
$\sigma$ is a random distance error expressed in term of a distance
modulus. In the present analysis, the measure of the distance is $V_c$
(equation (\ref{eqn:Mabs_de})). Accordingly, the random distance error
comes from a random error $\Delta V_c$, which is evaluated as a
square root of a quadratic sum of a line-of-sight component of a
random peculiar velocity of galaxies $\sigma_{rand}$ and an
observational error of the recession velocity $\Delta V_h$,

\begin{equation}
\Delta V_c=\left[\sigma_{rand}^2+(\Delta V_h)^2\right]^{\frac{1}{2}}\,.
\label{eqn:deltaV}
\end{equation}

\noindent
Using $\Delta V_c$, we obtain
$\sigma=5\log e\cdot(\Delta V_c/V_c^{(i)})$.
The systematic error $\delta$ caused by the {\em homogeneous}\/
Malmquist bias is therefore given by

\begin{equation}
\delta=15\log e\cdot\left(\frac{\Delta V_c}{V_c^{(i)}}\right)^2\,.
\label{eqn:Malm_bias}
\end{equation}

\noindent
An effect of a spatial {\em inhomogeneity}\/ of galaxy distribution is
discussed later in \S \ref{sec:MONTE-CARLO SIMULATIONS Examination of
the Present Method}.

We assume that an intrinsic broadening of the TF relation follows a
Gaussian function with a dispersion $\sigma_{T\!F}$. Then, using
$M_{T\!F}^{(i)}$ and $M_B^{(i)}$, we obtain the probability
${\cal P}^{(i)}$ in a following form\,:

\begin{equation}
{\cal P}^{(i)}=\frac{1}{C^{(i)}}\exp\left[-\frac{(M_{T\!F}^{(i)}-M_B^{(i)})^2}{2
\sigma_{(i)}^2}\right]\,,
\label{eqn:prob_TF}
\end{equation}

\noindent
where $\sigma_{(i)}$ includes $\sigma_{T\!F}$ and uncertainties of
$M_{T\!F}^{(i)}$ and of $M_B^{(i)}$, i.e.,

\begin{equation}
\sigma_{(i)}=\left[\sigma_{T\!F}^2+(\Delta M_{T\!F}^{(i)})^2+(\Delta M_B^{(i)})^
2\right]^{\frac{1}{2}}\,.
\label{eqn:sigma2}
\end{equation}

\noindent
The uncertainty $\Delta M_{T\!F}^{(i)}$ is evaluated from equations
(\ref{eqn:W0 calculation}), (\ref{eqn:R to sini}) and
(\ref{eqn:general TFrelation}) as

\begin{equation}
\Delta M_{T\!F}^{(i)} = \left[\left(\frac{\Delta W_{20}^{obs}}{W_{20}^{obs(i)}}\right)^2+\left(\frac{\Delta V_h}{c+V_h^{(i)}}\right)^2+\left(\frac{\ln 10\cdot\Delta\log R_{25}}{R_{25}^{(i)2}-1}\right)^2\right]^{\frac{1}{2}}\,,
\label{eqn:dMTF}
\end{equation}

\noindent
where $c$ is the velocity of light. The uncertainty
$\Delta M_B^{(i)}$, on the other hand, is computed from equation
(\ref{eqn:Mabs_de}) as

\begin{equation}
\Delta M_B^{(i)}=\left[(\Delta B_T^{0(i)})^2+\left(5\log e\cdot\frac{\Delta V_c}{V_c^{(i)}}\right)^2\right]^{\frac{1}{2}}\,.
\label{eqn:deltaMB}
\end{equation}

\noindent
With the selection criterion $2.30\leq\log W_{20}<2.75$ (\S
\ref{sec:DATA Sample selection}), the factor $C^{(i)}$ is computed by

\begin{equation}
C^{(i)}=\int_{M_{T\!F}^{l}}^{M_{T\!F}^{u}}\exp\left[-\frac{(x-M_B^{(i)})^2}{2\sigma_{(i)}^2}\right]dx\,,
\label{eqn:C}
\end{equation}

\noindent
where $M_{T\!F}^{l}$ and $M_{T\!F}^{u}$ are TF magnitudes corresponding
to the boundaries of $\log W_{20}=2.30$ and 2.75\,,
respectively.

The logarithmic likelihood function is finally obtained as

\begin{eqnarray}
\log{\cal L} & = & \log\prod_{i}{\cal P}^{(i)} \nonumber \\
	     & = & -\log e\cdot\sum_{i}\frac{(M_{T\!F}^{(i)}-M_B^{(i)})^2}{2\sigma _{(i)}^2}-\sum_{i}\log C^{(i)}\,.
\label{eqn:logL}
\end{eqnarray}

\noindent
The most probable values for parameters in question are determined
by maximizing $\log{\cal L}$.

We first apply the method to local calibrators to calibrate the TF
relation (\S \ref{sec:THE TULLY-FISHER RELATION Calibration}). In this
case, we let A, B in equation (\ref{eqn:general TFrelation}) and
$\sigma_{T\!F}$ in equation (\ref{eqn:sigma2}) be free parameters. The
method is then applied to our TF sample (\S \ref{sec:RESULT}) with the
calibrated TF relation to determine the unbiased $H_0$. In this case, we
let $h$ in equation (\ref{eqn:Mabs_de}) and $\sigma_{T\!F}$ in equation
(\ref{eqn:sigma2}) be free parameters.

There are five preset parameters to specify in our method;
$\Delta W_{20}^{obs}$, $\sigma_{rand}$, $\Delta V_h$,
$\Delta\log R_{25}$, and $\Delta B_T^0$. When the method is applied to
local calibrators, values for $\Delta W_{20}^{obs}$,
$\Delta\log R_{25}$, and $\Delta B_T^0$ are taken from RC3. The
parameters $\sigma_{rand}$ and $\Delta V_h$ are unnecessary in this case
because we do not use equation (\ref{eqn:Mabs_de}) but use a known
distance modulus to obtain $M_B$ for local calibrators. When the
method is applied to our TF sample, on the other hand, following
values are used in common to all the galaxies in the sample.
$\Delta W_{20}^{obs}=20\,$\kms , which is estimated as a square root
of a quadratic sum of a measurement error $\sim 10$\,\kms\ for
$W_{50}^{obs}$ (Giovanelli and Haynes 1985) and an uncertainty
$\sim 15$\,\kms\ associated with the line-profile width conversion by
equation (\ref{eqn:W50toW20}). $\sigma_{rand}=300\,$\kms , which is
taken from Davis and Peebles (1983), de Lapparent, Geller and Huchra
(1988) and Fisher et al. (1994).
$\Delta V_h=10\,$\kms\ is taken from Giovanelli and Haynes (1985).
$\Delta\log R_{25}=0.06$ is taken from our error assessment
(Watanabe 1996). $\Delta B_T^0=0.16\,$mag, which is
evaluated as a square root of a quadratic sum of our photometric
uncertainty $\Delta B_T=0.13\,$mag (Watanabe 1996) and
uncertainties $\Delta A_G=0.07\,$mag and
$\Delta A_i=0.05\,$mag which is evaluated from
$\Delta\log R_{25}$. Uncertainties of these values for the five
preset parameters are taken into account in an assessment of an
external error for $H_0$ (\S \ref{sec:RESULT Global Value}).

\section{THE TULLY-FISHER RELATION}
\label{sec:THE TULLY-FISHER RELATION}
\subsection{Calibration}
\label{sec:THE TULLY-FISHER RELATION Calibration}

The TF relation is calibrated using local calibrators whose distances
have been determined with Cepheid variables. These calibrators are
further restricted to those which meet the same selection criteria as
imposed on
the Pisces-Perseus galaxies, i.e., a morphological type Sa-Sd and an
inclination $i\geq 45^{\circ}$ (\S \ref{sec:DATA Sample selection}).
Ten spirals meet the criteria and are listed in Table
\ref{tbl:B-band local calibrators}. Among the ten calibrators, NGC\,224,
598, and 3031 were recently suspected to be unsuitable for the TF
calibration because of perturbations in the internal velocity field
(\cite{Giovanelli97a}). Further we find that NGC\,1365 and 4639 show
large difference in their inclination between RC3 and Giovanelli et al.
(1997a) ($i=59^{\circ}$ and $48^{\circ}$ for NGC\,1365 and
$i=49^{\circ}$ and $55^{\circ}$ for NGC\,4639). We therefore exclude
the five galaxies and use the remaining five as a fiducial set of
calibrators. Calibration using all the ten calibrators gives only slight
change in the following analysis. The alternative result is mentioned
when necessary.

Distance moduli $\mu^{(i)}$ of the calibrators are taken from the
literature (Table \ref{tbl:B-band local calibrators}). These distances
are all based on the LMC distance modulus of $18.5\,$mag
(e.g., Madore and Freedman 1991).
Other observational data are all taken from RC3. Using
$\mu^{(i)}$ and apparent magnitudes $B_T^{0(i)}$, absolute magnitudes
$M_B^{(i)}$ of the local calibrators are given by

\begin{equation}
M_B^{(i)}=B_T^{0(i)}-\mu^{(i)}\,.
\label{eqn:Mabs}
\end{equation}

\begin{figure}
\begin{center}
\epsfig{file=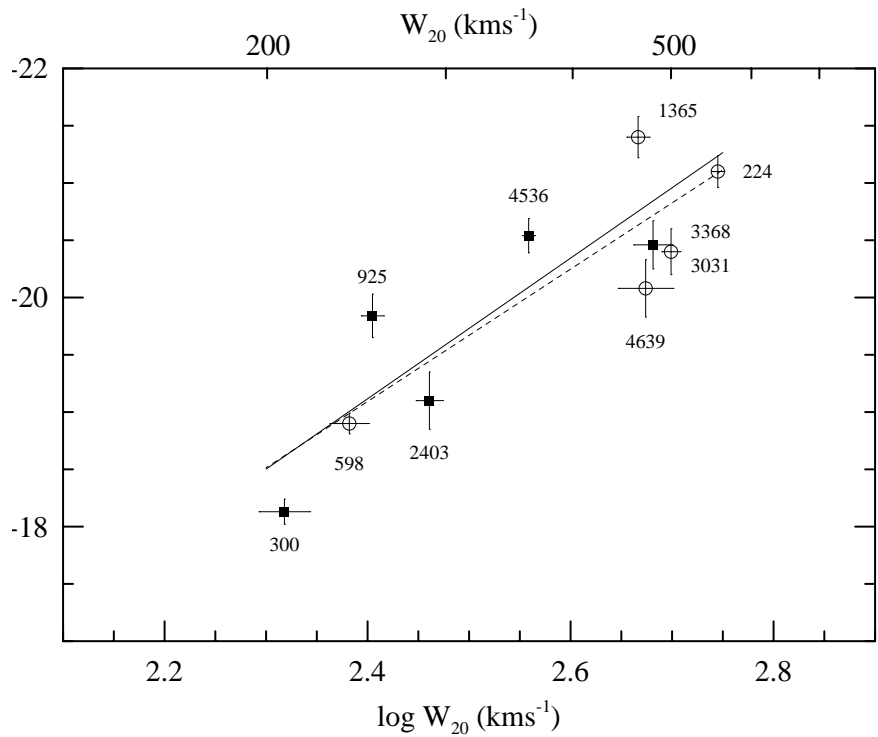,
        width=350pt, height=300pt, 
	bbllx=70pt, bburx=770pt, bblly=20pt, bbury=620pt, 
	clip=, angle=0}
\end{center}\vspace{-5mm}
\caption{The $B$-band Tully-Fisher diagram for the local calibrators
	 given in Table 2. The NGC number is labeled. Open circles
	 represent five galaxies discarded from the fiducial calibration
	 (see text). The solid and dashed lines show regression lines
	 for the fiducial set of five calibrators and for all the ten
	 calibrators, respectively. \label{fig:TFcalib_Fr}}
\end{figure}

A TF diagram of the calibrators is shown in Fig.\ref{fig:TFcalib_Fr}.
The range of $W_{20}$ covered by the local calibrators,
$2.30\le\log W_{20}\le 2.75$, is already adopted as one of the selection
criteria for the spiral sample (\S \ref{sec:DATA Sample selection}).
Letting A, B and $\sigma_{T\!F}$ be free parameters, we apply the
maximum likelihood method and then obtained the following calibration ;

\begin{equation}
M_{T\!F}^{(B)}=-19.73\,(\pm 0.20)-6.14\,(\pm 1.60)\,(\log W_{20}-2.5)\,.
\label{eqn:TF_B}
\end{equation}

\noindent
The scatter $\sigma_{T\!F}$ is 0.45\,mag. If we use all the ten
calibrators, we obtain A$=-19.67\pm 0.15$, B$=-5.78\pm 0.95$, and
$\sigma_{T\!F}=0.44$\,mag.
Our slopes are both consistent with that obtained by Strauss \& Willick
(1995) (B$=-5.3$) within the errors, while the values are marginally
smaller than that given by Pierce \& Tully (1992) (B$=-7.48$). This
discrepancy is ascribed to the lack of our calibrators with
$\log W_{20}<2.30$ that were used by them.

\subsection{Examination of the TF Magnitude of the Spiral Sample}
\label{sec:THE TULLY-FISHER RELATION Consistency between the Local
Calibrators and the TF Sample}

\begin{figure}
\begin{center}
\epsfig{file=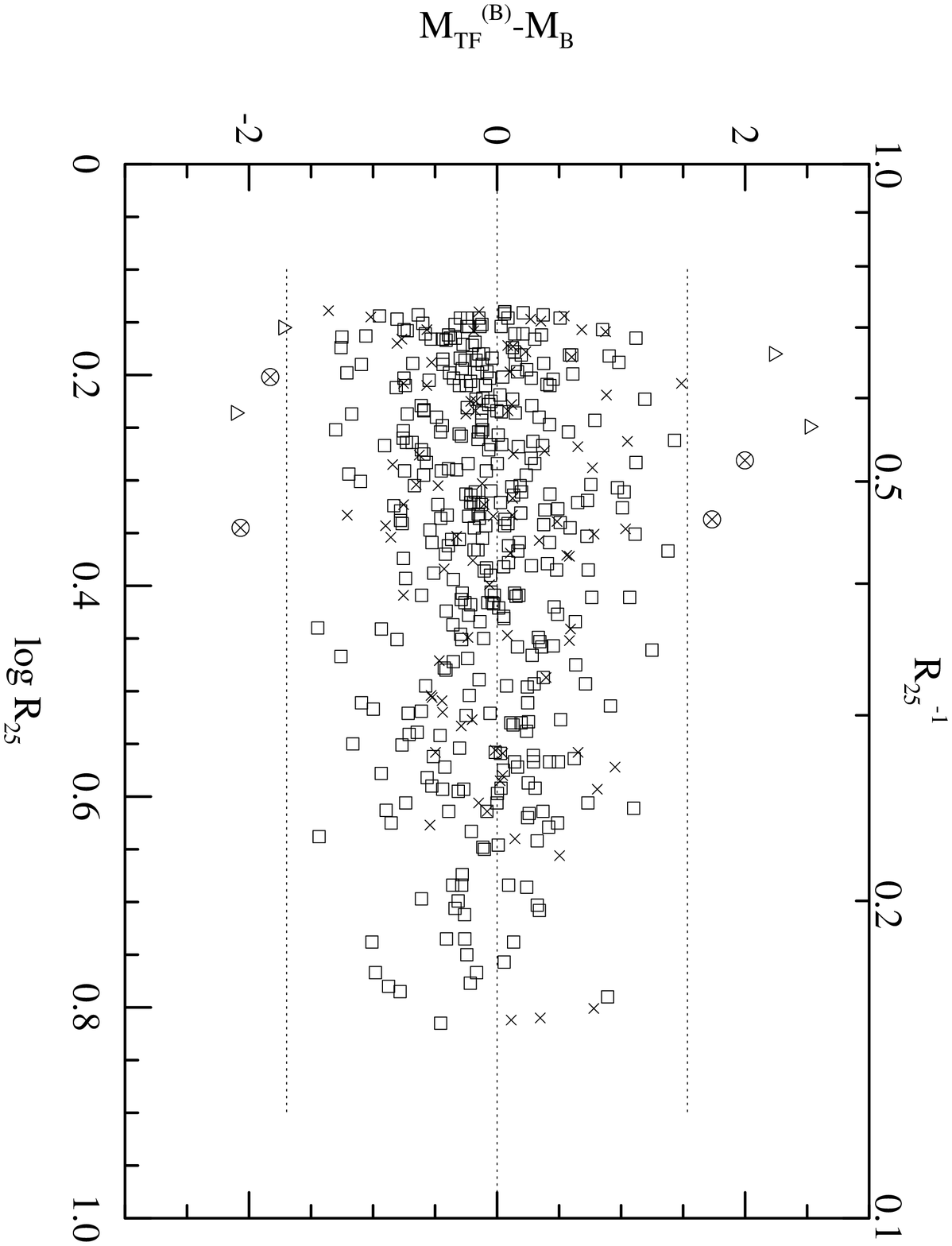,
        width=350pt, height=300pt,
        bbllx=70pt, bburx=770pt, bblly=20pt, bbury=620pt, 
        clip=, angle=0}
\end{center}\vspace{-5mm}
\caption{Difference between the TF magnitude $M_{T\!F}^{(B)}$ and the
	 magnitude $M_B$ from the Hubble's law with $\log h=-0.15$,
	 plotted against $\log R_{25}$ for the spiral sample
	 ($N=449$). Crosses represent galaxies whose images are
         contaminated by other images. Upper and lower dashed lines
	 delineate the boundaries $|M_{T\!F}^{(B)}-M_B|=1.62$\,mag.
	 Eight galaxies (encircled crosses and triangles) are out of the
	 boundaries and hence discarded from the TF analysis. The
	 enclosed crosses indicate galaxies with unreliable
	 observational data (see text).
	 \label{fig:logR_vs_dMabs_449}}
\end{figure}

\begin{figure}
\vspace*{-15mm}
\begin{center}
\epsfig{file=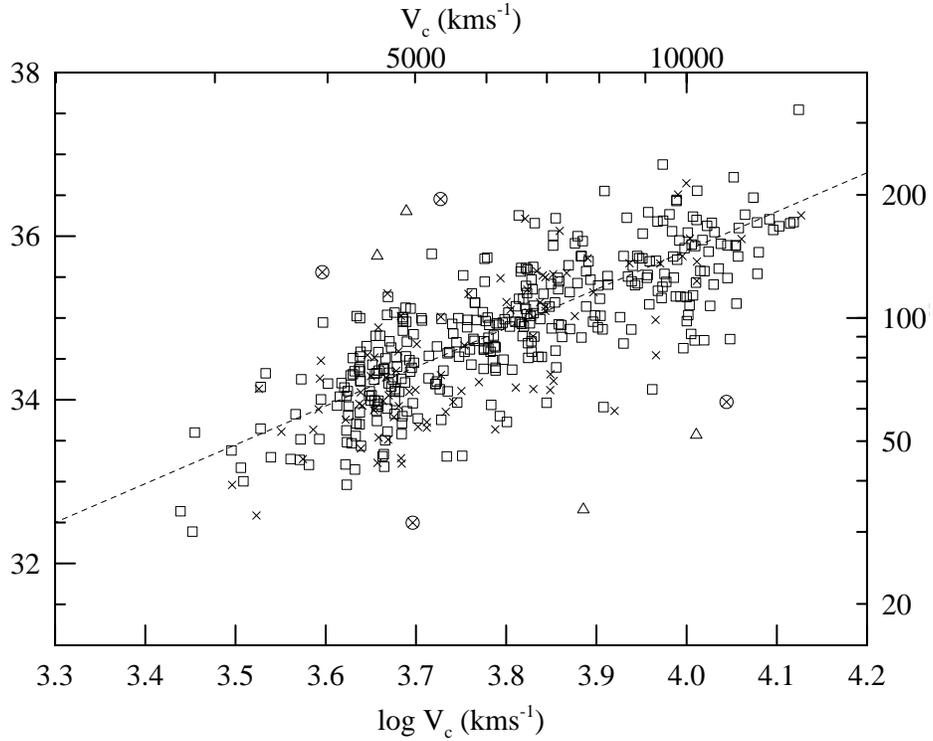,
        width=350pt, height=300pt, 
	bbllx=70pt, bburx=770pt, bblly=20pt, bbury=620pt, 
	clip=, angle=0}
\end{center}\vspace{-5mm}
\caption{The velocity-TF distance relation for the spiral sample
         ($N=449$). Symbols are the same as in Fig.6.
	 A dashed line represents an arbitrary Hubble's law with
	 $\log h=-0.15$. \label{fig:logV_vs_DM_449}}
\end{figure}

In determination of $M_{T\!F}^{(B)}$, the inclination $i$ is the only
observable derived from our measurements (equation
[\ref{eqn:R to sini}]). To examine an error in $M_{T\!F}^{(B)}$ caused
by an error of our $R_{25}$, we plot in Fig.\ref{fig:logR_vs_dMabs_449}
the residual $M_{T\!F}^{(B)}-M_B$ as a function of $\log R_{25}$. The
absolute magnitude $M_B$ is computed from the Hubble's law (equation
(\ref{eqn:Mabs_de})) with $H_0$ arbitrarily assumed to be
$\log h=-0.15$ ($H_0=70.8\,$\kmsMpc ). Accordingly, a global offset
of the plot from a horizontal line of $M_{T\!F}^{(B)}-M_B=0$ is of
little significance. We
found no significant systematic trend for $M_{T\!F}^{(B)}-M_B$ as a
function of $\log R_{25}$ but found several deviant galaxies. Dispersion
of the plot is
$0.53$\,mag
and, if we assume a Gaussian
distribution for $M_{T\!F}^{(B)}-M_B$ for 449 galaxies, less than 0.5
galaxies fall in the range out of
$3.06\sigma$.
Accordingly, deviation larger than
$|M_{T\!F}^{(B)}-M_B|=1.62$\,mag is considered to be significant, and
we find eight galaxies out of the boundaries. Four out of the eight have
possible reasons for the deviation as follows. PGC\,6802 has very low
surface brightness and then unreliable $B_T$ and $R_{25}$. PGC\,10013
has a very wavy growth curve which led unreliable $B_T$.
PGC\,10740 is extremely faint ($B_T = 17.09$\,mag) and hence its
photometry may be unreliable. PGC\,14033 suffers from severe
contamination of HI flux of nearby galaxies (\cite{GH85}). We discard
these from the TF analysis. Although we
found no reason for the large deviation to the remaining four
galaxies, we further exclude them from the TF analysis
and hereafter refer to the remaining 441 galaxies as the TF sample
(Table \ref{tbl:Sample selection for the TF analysis}). An
effect of including the latter four galaxies is mentioned when
necessary. The velocity-distance relation
is shown in
Fig.\ref{fig:logV_vs_DM_449}, where global agreement with the Hubble's
law is evident.

\section{MONTE-CARLO SIMULATIONS} \label{sec:MONTE-CARLO SIMULATIONS}
\subsection{Simulated Sample}
\label{sec:MONTE-CARLO SIMULATIONS Description of the Simulated sample}

Evaluation of the effect of biases is an important issue in this study.
However, it is quite difficult to evaluate the effect analytically
because physical properties, spatial distribution, and observational
limits for galaxies are entangled so complicatedly. For this purpose,
a Monte-Carlo simulation provides a straightforward and convincing test
(e.g., Kolatt et al. 1996).

We perform Monte-Carlo simulations using pseudo samples of 441 galaxies.
These samples are made with unknown parameters given a priori as
$\log h=-0.15$
($H_0=70.8$\,\kmsMpc ) and $\sigma_{T\!F}=0.40\,$mag. Procedure of
assigning observational parameters to individual galaxies is described
below.

Galaxies are distributed uniformly with respect to a pure Hubble flow
velocity $V_H$. In the range of
$4000\leq V_H<5000\,$\kms , five times as many galaxies as
those in the uniform background are added to imitate the clustering
feature of the \PP\ supercluster. Galaxies in
$1000\leq V_H<4000\,$\kms\ are randomly discarded so that the
number density becomes one third of that of the uniform background so as
to model the low number density space in the foreground of the
supercluster (Haynes and Giovanelli 1986). A true distance is given to
the galaxies via the Hubble's law with $V_H$ and $\log h=-0.15$. We
then define a velocity $V_c$ by adding to $V_H$ a Gaussian random
motion with dispersion of $\sigma_{rand}=300\,$\kms\
(\cite{DP83,Lapparent88,Fisher94}). Note that observers know neither
$V_H$ nor the true distance, but only $V_c$ by observations as described
later in this section.

Absolute magnitudes are assigned to the galaxies according to a galaxian
luminosity function. Since our maximum likelihood analysis is little
affected by the luminosity function, it is unnecessary to pursue its
true (unknown) shape and amplitude. We adopted a Schechter form with
$M_B^\ast=-20.0$\,mag and $\alpha=-1.1$, which are representative values
found in recent studies (Efstathiou, Ellis and Peterson 1988; Loveday
et al. 1992; Marzke, Huchra and Geller 1994).
Then, an intrinsic HI line-profile width $W$ is computed from the
absolute magnitude and the TF relation given by equation
(\ref{eqn:TF_B}). It is assumed that the TF relation is broadened by
its own intrinsic scatter with a Gaussian probability function of
$\sigma_{T\!F}=0.40\,$mag. Inclination $i$ is
assigned so that the galaxies have random orientation. Using $W$, $i$,
and $V_c$, a projected line-profile width $W^{obs}$ is obtained via
equation (\ref{eqn:W0 calculation}). Given the intrinsic axial ratio
$R_0^{-1}=0.2\,$, an axial ratio $R$ is derived from the
inclination using equation (\ref{eqn:R to sini}). An apparent magnitude
$B_T^0$ is derived from the absolute magnitude and the true distance.
Internal extinction and K-dimming are computed according to a
prescription given in RC3.

\begin{figure}
\vspace{-10mm}
\begin{center}
\epsfig{file=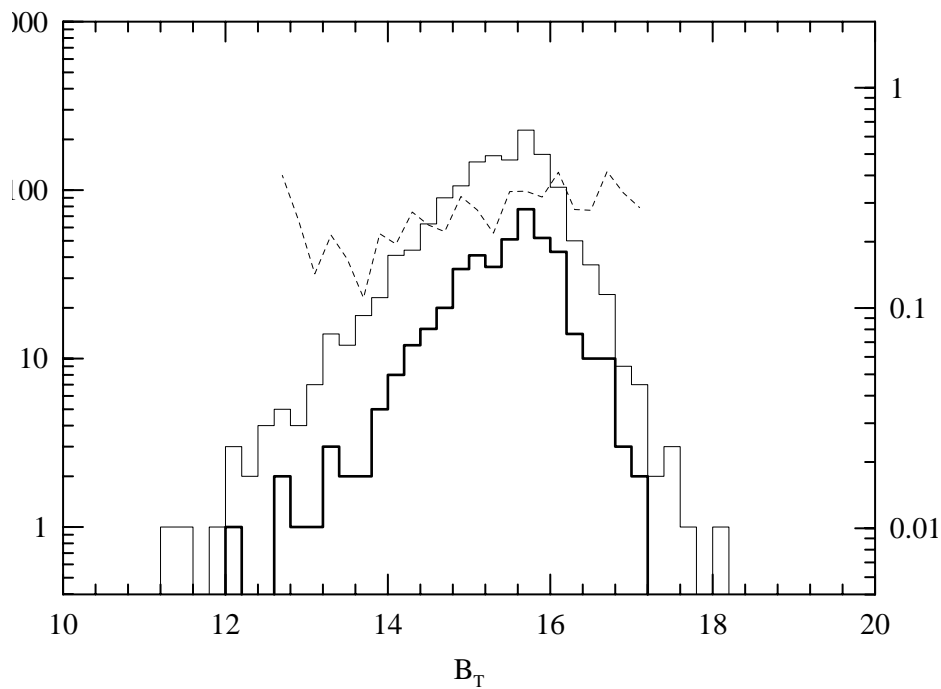,
        width=350pt, height=300pt,
        bbllx=70pt, bburx=770pt, bblly=20pt, bbury=620pt, 
        clip=, angle=0}
\end{center}\vspace{-5mm}
\caption{A histogram of the total magnitude $B_T$ for the spiral
         sample ($N=449$, thick line) and the photometry sample
	 ($N=1524$, thin line). A dashed line shows a number
	 ratio of the former sample to the latter.
	 \label{fig:Bt_vs_N_1524_957_449}}
\end{figure}

Now, these simulated galaxies are {\em observed}\/. The virtual
observation imposes observational errors to the parameters and to reduce
the sample according to a limiting magnitude. First, observational
errors are added to $V_c$, $W^{obs}$, $R$, and $B_T^0$. The errors are
assumed to follow random Gaussians whose dispersions are estimated from
the actual TF sample galaxies. Following values are used for the
dispersions \,: $\Delta V=10\,$\kms\,,
$\Delta W^{obs}=20\,$\kms\,, $\Delta\log R=0.06$\,, and
$\Delta B_T^0=0.16\,$mag (\S \ref{sec:A NEW METHOD OF THE ANALYSIS
Formalism}). The simulated galaxies are then selected according to a
limiting magnitude. A selection function ${\cal S}(B_T)$ is defined as

\begin{equation}
{\cal S}(B_T)=\left\{
  \begin{array}{ll}
  1 & (B_T<15.0\,\mbox{mag}) \\
  N(B_T)/N_0 10^{0.6B_T} & (15.0\le B_T<17.0\,\mbox{mag}) \\
  0 & (B_T\ge 17.0\,\mbox{mag})\,,
  \end{array}
  \right.
\label{eqn:selection_function}
\end{equation}

\noindent
where $N(B_T)$ is the number of galaxies in the actual TF sample as a
function of $B_T$ (Fig.\ref{fig:Bt_vs_N_1524_957_449}). The
normalization factor $N_0$ is determined by a least-squares fit of
$N=N_0\cdot 10^{0.6B_T}$ to the number distribution of the subsample of
$B_T<15.0$\,mag shown in Fig.\ref{fig:Bt_vs_N_1524_957_449}.

\begin{figure}
\begin{center}
\epsfig{file=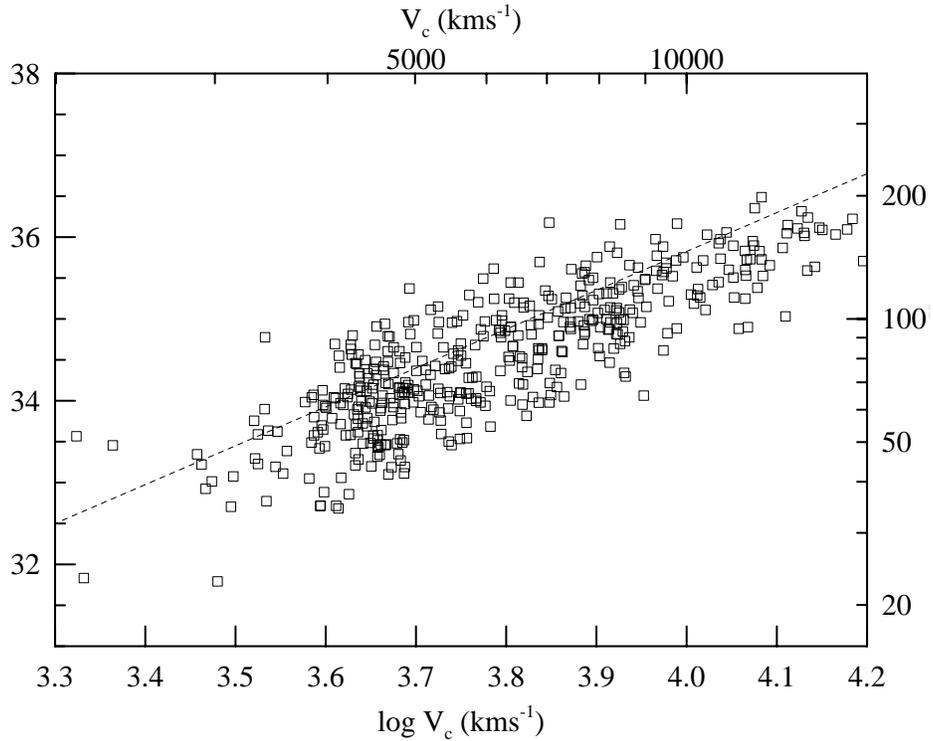,
        width=350pt, height=300pt, 
	bbllx=70pt, bburx=770pt, bblly=20pt, bbury=620pt, 
	clip=, angle=0}
\end{center} \vspace{-5mm}
\caption{Same as Fig.7 but for one of the simulated TF
         samples ($N=441$). \label{fig:logV_vs_DM_Bclu}}
\end{figure}

Finally, according to the same selection criteria as imposed on the
actual TF sample, i.e., $V_c\geq 1000\,$\kms , $i\geq 45^{\circ}$
and $2.30\leq\log W_{20}<2.75\,$, the simulated magnitude-limited
sample is reduced to generate a simulated TF sample of 441 galaxies.
A velocity-TF distance relation of a typical example of the simulated TF
samples is shown in Fig.\ref{fig:logV_vs_DM_Bclu}.

\subsection{Examination of the Performance of the Present Method}
\label{sec:MONTE-CARLO SIMULATIONS Examination of the Present Method}

\begin{figure}
\vspace*{-25mm}
\begin{center}
\epsfig{file=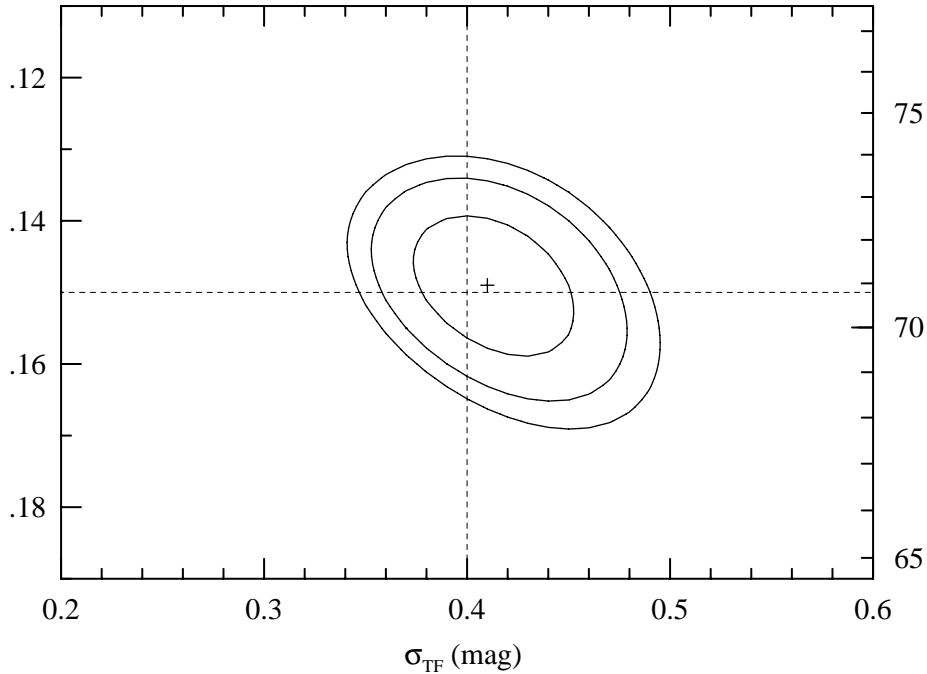,
        width=350pt, height=300pt, 
	bbllx=70pt, bburx=770pt, bblly=20pt, bbury=620pt, 
	clip=, angle=0}
\end{center}\vspace{-5mm}
\caption{Probability contour map for $H_0$ and $\sigma_{T\!F}$
	 obtained from one of the simulated TF samples. Confidence
	 levels at 70\%, 95\% and 99\% are shown. Horizontal and
	 vertical dashed lines indicate the input values $\log h=-0.15$
	 and $\sigma_{T\!F}=0.40$\,mag, respectively, which are assumed
	 a priori for the simulated sample.
	 \label{fig:sTF_vs_logh_Bclu}}
\end{figure}

\begin{figure}
\begin{center}\vspace{-20mm}
\epsfig{file=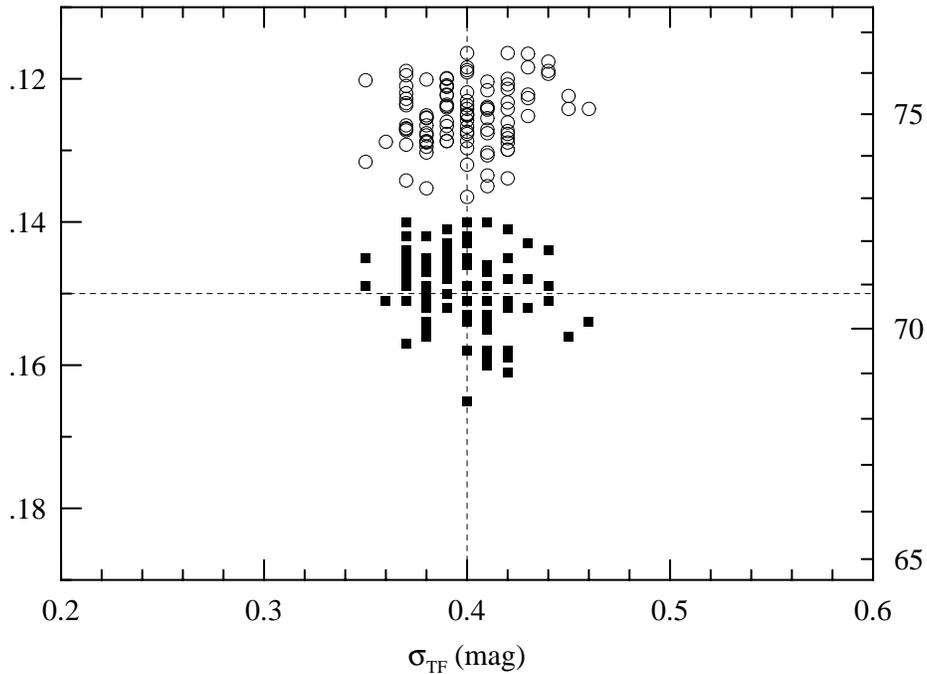,
        width=350pt, height=300pt, 
	bbllx=70pt, bburx=770pt, bblly=20pt, bbury=620pt, 
	clip=, angle=0}
\end{center}\vspace{-5mm}
\caption{The most probable values of $H_0$ and $\sigma_{T\!F}$
	 for one hundred different simulated TF samples. Filled squares
	 represent those obtained by the maximum likelihood method,
	 while open circles represent sample averages of
	 $\log(V_c/r_{T\!F})$ for $\log h$ with $\sigma_{T\!F}$
	 a priori set equal to that obtained with the maximum
	 likelihood method. \label{fig:sTF_vs_logh_Bclu_external}}
\end{figure}

Our TF analysis with the maximum likelihood method is applied to one
hundred different simulated TF samples. We show in
Fig.\ref{fig:sTF_vs_logh_Bclu} a probability contour map for $H_0$ and
$\sigma_{T\!F}$ obtained for one of the simulated samples. The contours
represent only an internal error associated with the likelihood
calculation. The input values $\log h=-0.15$ and
$\sigma_{T\!F}=0.40$\,mag are reproduced within the internal
error of the maximum likelihood method (70\% confidence level). A 
distribution of the most probable values of
($H_0$,$\sigma_{T\!F}$) is shown in
Fig.\ref{fig:sTF_vs_logh_Bclu_external} for the one hundred simulated
samples. Also shown is a distribution of sample averages of
$\log (V_c/r_{T\!F})$ for the same samples. In this case, the intrinsic
scatter $\sigma_{T\!F}$ is tentatively taken equal to the value obtained
by the maximum likelihood method. It is clearly demonstrated that the
sample average of $\log (V_c/r_{T\!F})$ is subject to a significant
systematic error. The $H_0$'s derived from the maximum likelihood
method, on the
other hand, show a distribution consistent with the input value.

\begin{figure}
\vspace*{-15mm}
\begin{center}
\epsfig{file=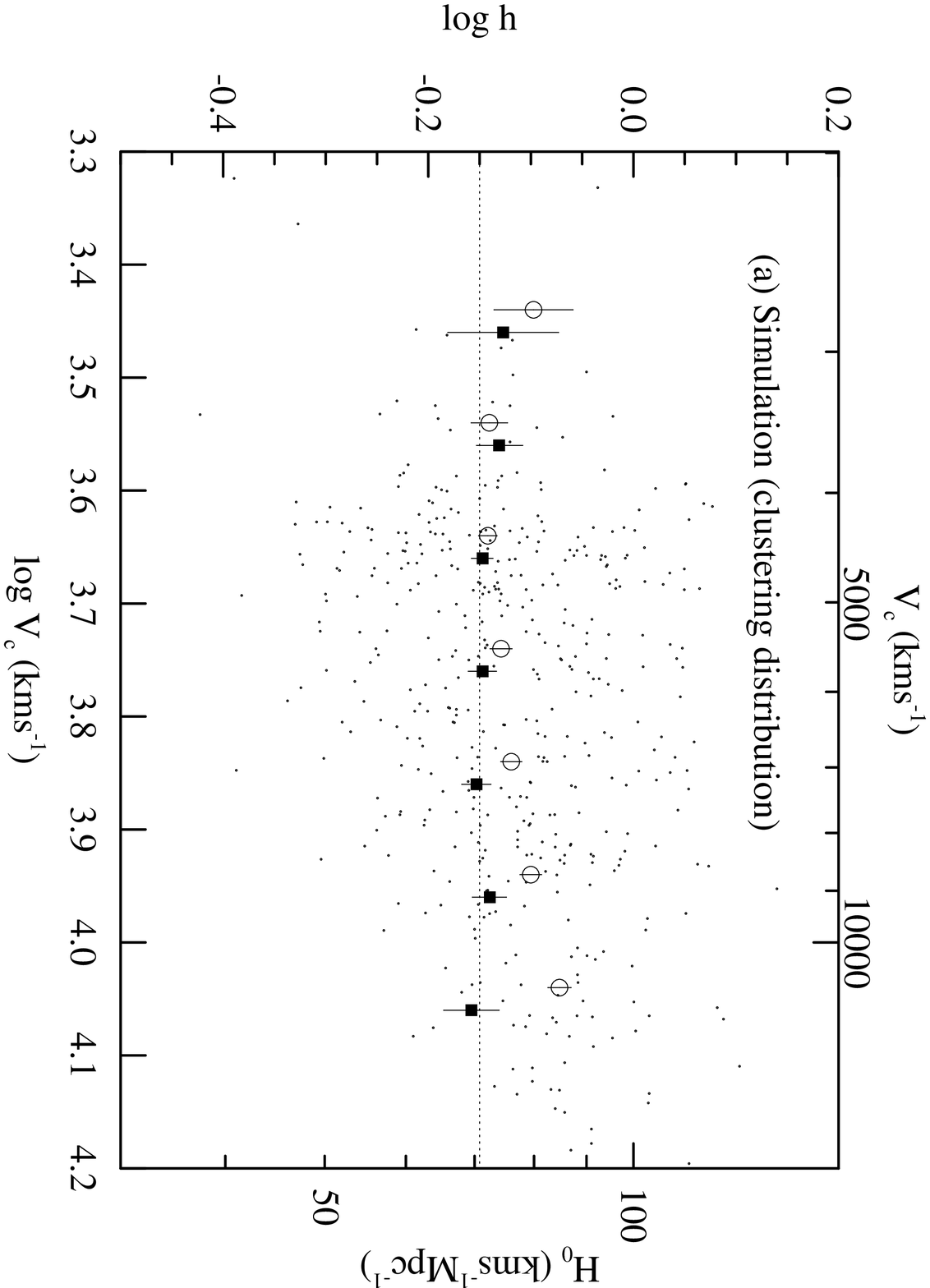,
        width=350pt, height=300pt, rheight=280pt,
        bbllx=70pt, bburx=770pt, bblly=20pt, bbury=620pt, 
        clip=, angle=0}\vspace{5mm}
\epsfig{file=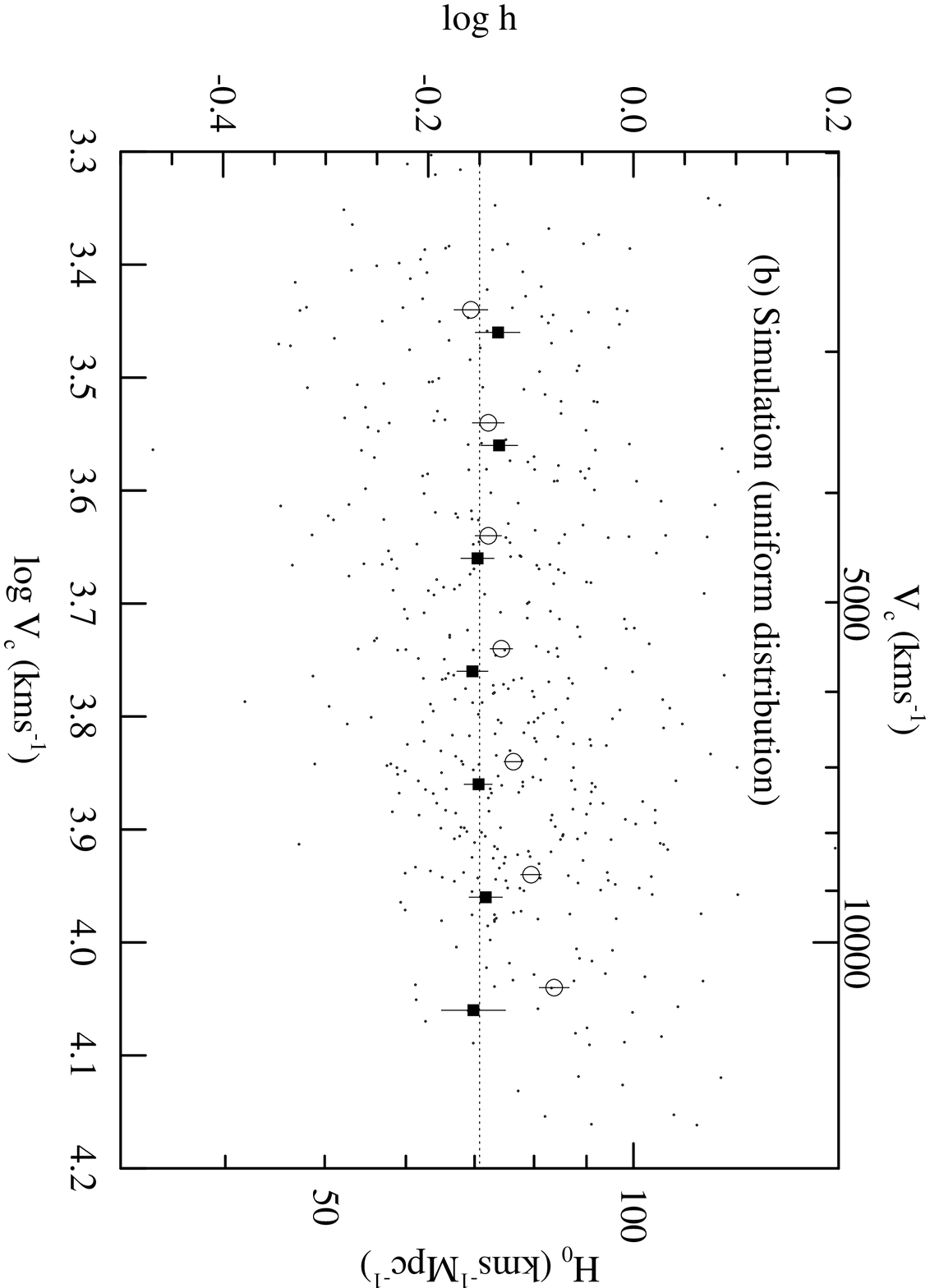,
        width=350pt, height=300pt, 
	bbllx=70pt, bburx=770pt, bblly=20pt, bbury=620pt, 
	clip=, angle=0}
\end{center} \vspace{-5mm}
\caption{Spatial variation of $H_0$ obtained ($a$) from the simulated TF
         sample shown in Fig.9
         and ($b$) from a simulated TF sample with an uniform spatial
	 distribution. The intrinsic scatter $\sigma_{T\!F}$ is fixed at
	 0.41\,mag for all the subsamples in $\log V_c$ bins. Dots
	 represent the individual galaxies, to which
	 the Hubble ratio $V_c/r_{T\!F}$ is given as $H_0$. Filled
	 squares represent the most probable $H_0$ in each bin obtained
	 by the maximum likelihood method, while open circles represent
	 the sample average of $\log(V_c/r_{T\!F})$. The squares and the
	 circles are shifted from the center of each $\log V_c$ bin to
	 avoid the overlap. Error bars are taken from the 70\% error
	 obtained for each subsample. A dashed line indicates the input
	 value $\log h=-0.15$. \label{fig:logV_vs_logh_Bsmo}}
\end{figure}

We further examine the performance of our method by computing a spatial
variation of $H_0$ with a simulated TF sample divided into bins of
$\log V_c$\,. The intrinsic scatter $\sigma_{T\!F}$ is fixed for all the
subsamples in the $\log V_c$ bins at 0.41\,mag obtained from the
analysis shown in Fig.\ref{fig:sTF_vs_logh_Bclu}. The variation of $H_0$
is shown in Fig.\ref{fig:logV_vs_logh_Bsmo}$a$.
The evident increase of the sample average of
$\log (V_c/r_{T\!F})$ with $\log V_c$ demonstrates the effect of the
sampling incompleteness bias. The apparent increase manifests itself as
the systematic error shown in Fig.\ref{fig:sTF_vs_logh_Bclu_external}.
It is evident, on the other hand, that the maximum likelihood method is
free from the systematic error caused by the sampling incompleteness
bias and gives unbiased estimates of $H_0$'s regardless of $\log V_c$.

As shown in equation (\ref{eqn:Mabs_de}), we corrected for the Malmquist
bias assuming an uniform galaxy distribution although the actual
distribution is inhomogeneous. To examine the effect of ignoring the
inhomogeneity, we made a comparative simulation using a uniform galaxy
distribution. A resultant variation of $H_0$ is shown in
Fig.\ref{fig:logV_vs_logh_Bsmo}$b$. Comparison of the two results would
reveal the effect of inhomogeneity, if any, as a residual. However, we
observe no significant difference between the two results.
Accordingly, we neglect the correction for the inhomogeneous Malmquist
bias in the present analysis.

\section{RESULT} \label{sec:RESULT}
\subsection{Global Value of $H_0$} \label{sec:RESULT Global Value}

\begin{figure}
\begin{center}
\epsfig{file=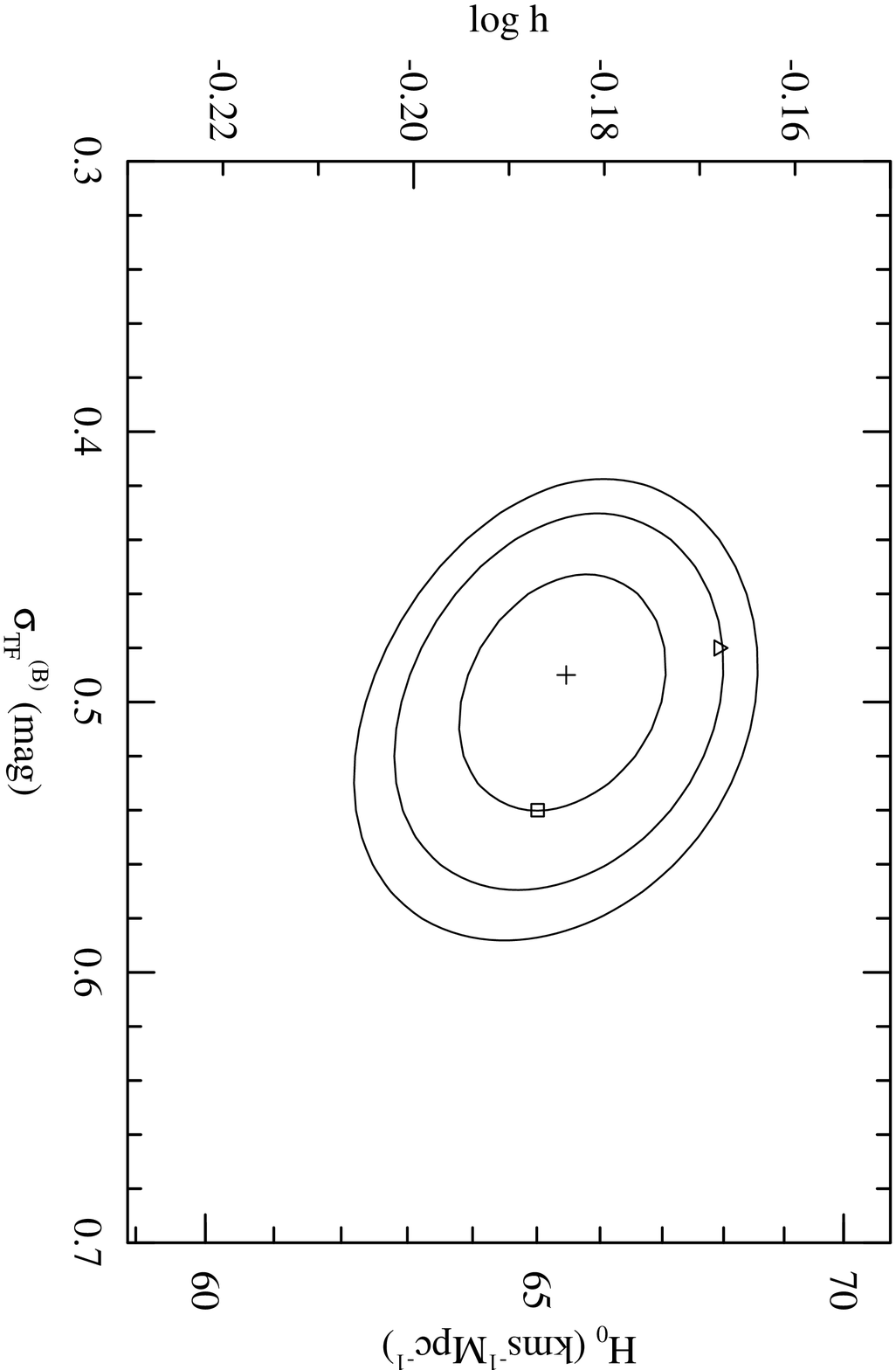,
        width=350pt, height=300pt, 
	bbllx=70pt, bburx=770pt, bblly=20pt, bbury=620pt, 
	clip=, angle=0}
\end{center}\vspace{-5mm}
\caption{Probability contour map for $H_0$ and $\sigma_{T\!F}^{(B)}$
         for the TF sample of 441 galaxies.
	 A plus indicates the most probable values. Contours are drawn
	 at 70\%, 95\% and 99\% confidence levels. A triangle and a
	 square indicate the results obtained from the ten calibrators
	 and from the inclusion of four deviant galaxies, respectively
	 (see text).  \label{fig:sTF_vs_logh_B441}}
\end{figure}

The maximum likelihood method is applied to our TF sample. The resultant
probability contour map for $H_0$ and $\sigma_{T\!F}^{(B)}$ is shown in
Fig.\ref{fig:sTF_vs_logh_B441}. Note that the contours
represents only the internal error of the likelihood calculation. Other
error sources are as follows. A zero point uncertainty of the TF
calibration is evaluated from the standard deviation of the calibrators
around the TF relation, i.e.,
$0.45/\sqrt{5}=0.20\,$mag
(\S \ref{sec:THE TULLY-FISHER RELATION Calibration}).
The systematic difference
of $A_G$ is examined among the schemes of BH, RC2, and RSA (Sandage and
Tammann 1981), and the difference of $A_i$ is examined among RC3, RC2,
RSA and Bothun et al. (1985). The result is summarized in Table
\ref{tbl:correction difference}.
Since the same
absorption correction schemes are used both for the TF sample and for
the local calibrators, scheme dependence tends to cancel out.
The maximum differences in $A_G$ and in
$A_i$ are adopted as the respective external errors.
These external systematic errors are summarized in Table
\ref{tbl:error summary}.

Finally, we obtain the best estimate of the unbiased global $H_0$ as

\begin{equation}
H_0=65\pm 2^{\;+20}_{\;-14}\,\mbox{\kmsMpc}\,.
\label{eqn:H0_400}
\end{equation}

\noindent
The first and the second terms of the quoted errors represent the
internal random error of the maximum likelihood method and the external
systematic error, respectively. Note that the latter is given by the
uppermost and the lowermost boundaries of the {\em linear}\/ sum.
The errors of $H_0$ are often underestimated by ignoring various
complicated systematic effects. Our error estimate is the result of our
best effort to include all the known error sources with conservative
error budgets assigned to respective sources.
The intrinsic scatter of the $B$-band TF relation is obtained to be
$\sigma_{T\!F}^{(B)}=0.49\pm 0.05$\,mag. An analysis with the ten
calibrators results in $H_0=68$\,\kmsMpc\ with
$\sigma_{T\!F}^{(B)}=0.48$\,mag. If we use 445 galaxies including
four spirals discarded from the TF sample (\S\ref{sec:THE TULLY-FISHER
RELATION Consistency between the Local Calibrators and the TF Sample}),
$\sigma_{T\!F}^{(B)}$ slightly increases to $0.54$\,mag but $H_0$ does
not change in practice.

\subsection{Spatial Variation of $H_0$}
\label{sec:RESULT Spatial Variation}

\begin{figure}
\begin{center}
\epsfig{file=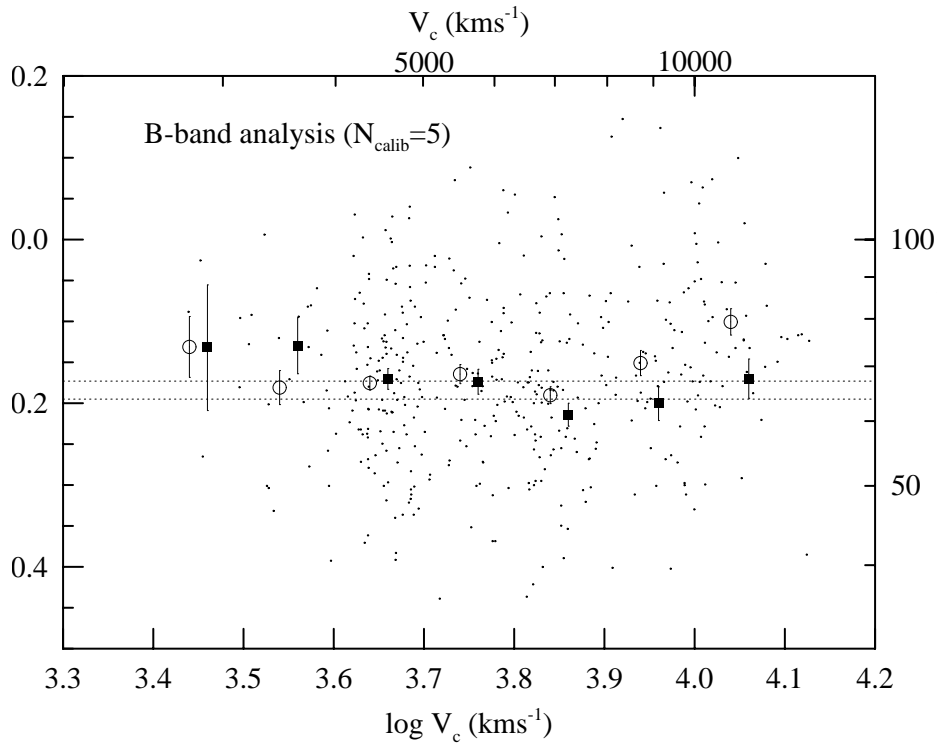,
        width=350pt, height=300pt, 
	bbllx=70pt, bburx=770pt, bblly=20pt, bbury=620pt, 
	clip=, angle=0}
\end{center}\vspace{-5mm}
\caption{Spatial variation of $H_0$ obtained from the TF sample.
	 The intrinsic scatter $\sigma_{T\!F}^{(B)}$ is fixed at
	 0.49\,mag for all the subsamples in $\log V_c$ bins. Dots
	 represent the individual galaxies, to which
	 the Hubble ratio $V_c/r_{T\!F}^{(B)}$ is given as $H_0$. Filled
	 squares represent the most probable $H_0$ in each bin obtained
	 by the maximum likelihood method, while open circles represent
	 the sample average of $\log(V_c/r_{T\!F}^{(B)})$. The squares
	 and
	 the circles are shifted from the center of each $\log V_c$ bin
	 to avoid the overlap. Error bars are taken from the 70\%
	 error obtained for each subsample. Two dashed lines indicate
	 the 70\% confidence levels $H_0=63$ and 67\,\kmsMpc\ shown
	 in Fig.12.  \label{fig:logV_vs_logh_B441}}
\end{figure}

A spatial variation of $H_0$ is examined using the TF sample divided
into bins of $\log V_c$. Dispersion $\sigma_{T\!F}^{(B)}$ is fixed at
0.49\,mag for all the subsamples in the bins; the free parameter is only
$H_0$ in the maximum likelihood method here. The spatial variation
obtained is shown in Fig.\ref{fig:logV_vs_logh_B441}. It is clearly
demonstrated that the sample means of $\log(V_c/r_{T\!F}^{(B)})$ are
seriously
subject to the systematic error as $V_c$ increases. On the other hand,
the amplitude of the variation of $H_0$'s is consistent with the global
$H_0$ within the errors except for poor subsamples at $\log V_c<3.6$.

\section{ANALYSIS WITH WILLICK'S r-BAND DATA}
\label{sec:ANALYSIS WITH WILLICK'S r-BAND SAMPLE}
\subsection{$r$-band Sample}
\label{sec:ANALYSIS WITH WILLICK'S r-BAND SAMPLE r-band sample}

\begin{figure}
\vspace*{-15mm}
\begin{center}
\epsfig{file=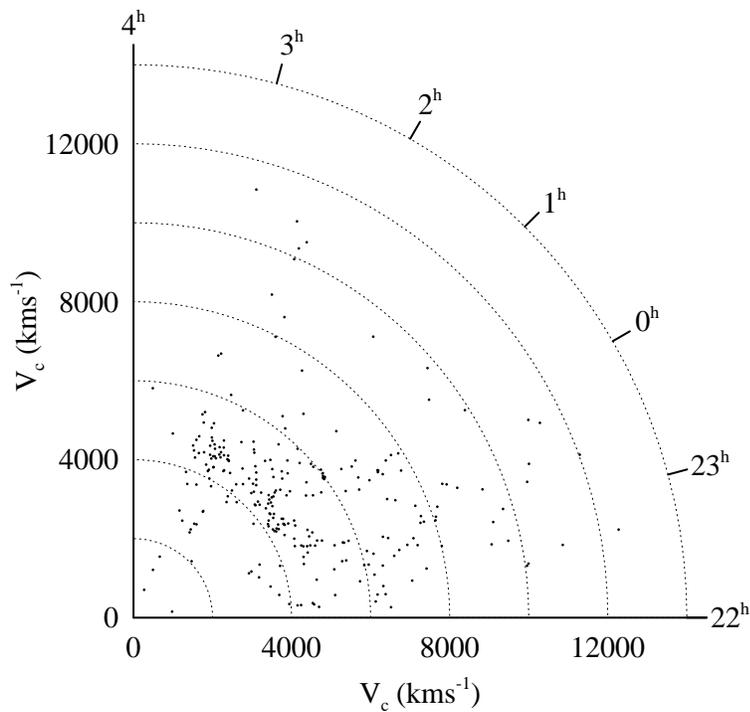,
        width=350pt, height=300pt, 
	bbllx=70pt, bburx=770pt, bblly=20pt, bbury=620pt, 
	clip=, angle=0}
\end{center}\vspace{-5mm}
\caption{Velocity distribution of the $r$-band TF sample ($N=271$).
	 The velocity is expressed in the CMB-rest frame.
         \label{fig:cz_plot_WilcomKiso_222}}
\end{figure}

The same $H_0$ determination is carried out using $r$-band CCD data of
the \PP\ region given by
Willick et al. (1997).
The sample (w91pp) is based on UGC and consists of
326 spirals complete
down to $m_z=14.5$\,mag. Application of the same selection criteria as
used in our TF analysis to w91pp reduced the sample to 271 galaxies.
We hereafter refer to this sample of 271 galaxies
as the $r$-band TF sample. The
velocity distribution is shown in Fig.\ref{fig:cz_plot_WilcomKiso_222}.
The shallower limiting magnitude than ours ($m_z=15.5$\,mag) makes
the spatial coverage of this sample less extensive than that of the
$B$-band sample (Fig.\ref{fig:cz_plot_449}).

\subsection{$r$-band TF calibration}
\label{sec:ANALYSIS WITH WILLICK'S r-BAND SAMPLE r-band TF calib}

The $r$-band TF relation was calibrated with six local calibrators
(Table \ref{tbl:B-band local calibrators}). The same criteria as in
$B$-band leave only two local calibrators (NGC\,1365 and 2403). We do
not exclude NGC\,224, 300, 598 and 3031 simply because the two
calibrators are insufficient for reasonable calibration.
Photometry data for NGC\,1365 are taken from
ESO-Uppsala catalog (\cite{ESO-Uppsala}) while the data for the others
are from Pierce and Tully (1992).
Kron-Cousins $R$ and $I$ magnitudes
are transformed with a color equation
$r=0.40+R-0.28\,(R-I)$ given by Willick (1991).
Total magnitudes $r_T$ were corrected for $A_G$ and $A_i$ using
BH and Bothun et al. (1985), respectively, with a transformation
$A^{(r)}=0.60A^{(B)}$.
Corrected apparent magnitudes $r_T^0$ are given in Table
\ref{tbl:B-band local calibrators}. With the maximum likelihood method,
the $r$-band TF relation is obtained as

\begin{equation}
M^{(r)}_{T\!F}=-20.15\,(\pm 0.18)-7.93\,(\pm 1.04)\,(\log W_{20}-2.5)\,.
\label{eqn:TF_r}
\end{equation}

\begin{figure}
\begin{center}
\epsfig{file=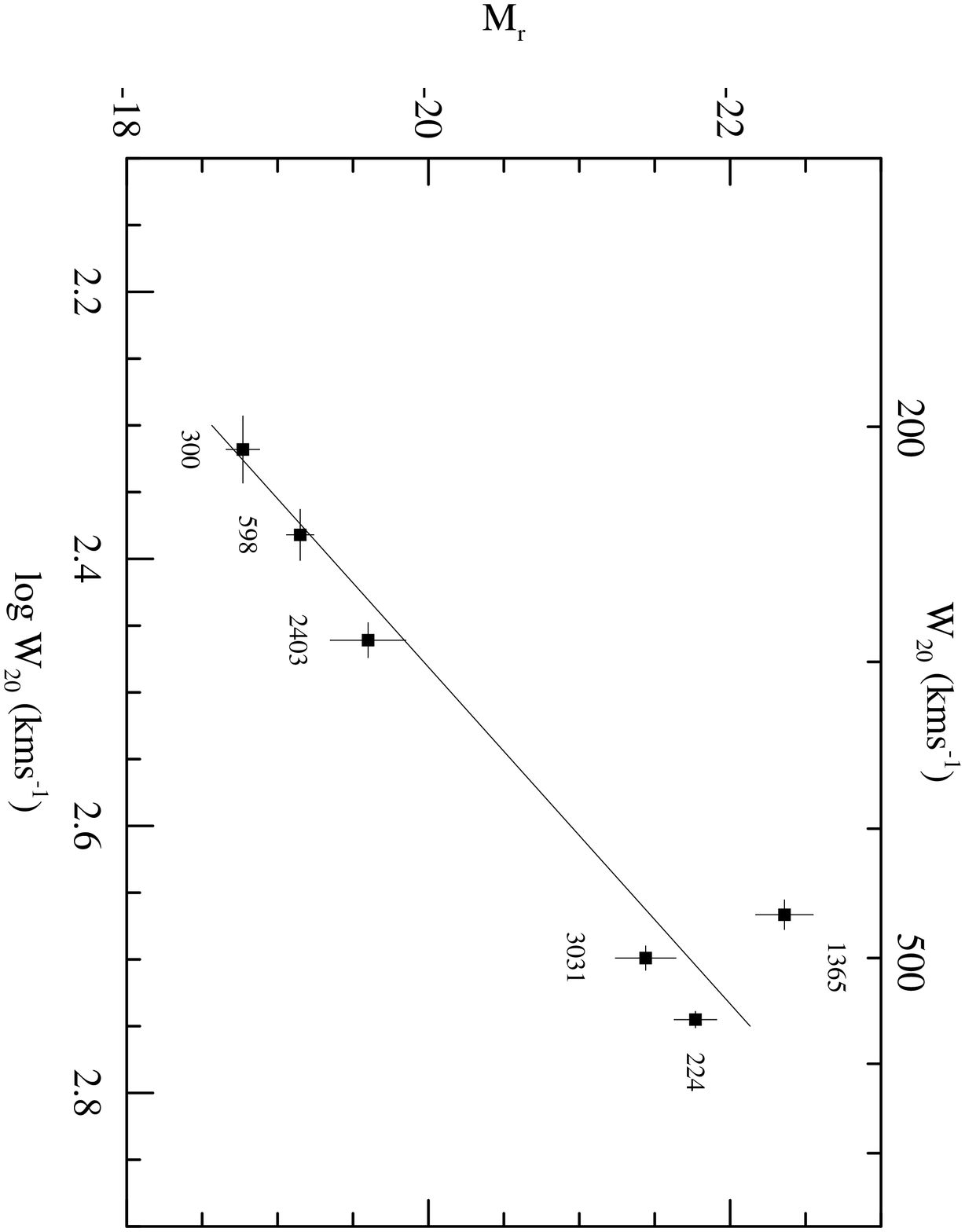,
        width=350pt, height=300pt, 
	bbllx=70pt, bburx=770pt, bblly=20pt, bbury=620pt, 
	clip=, angle=0}
\end{center}\vspace{-5mm}
\caption{The $r$-band \TF\ diagram for the local calibrators given in
         Table 2. The NGC number is labeled. The regression line shows
	 equation (18).
	 \label{fig:TFcalib_PT_r}}
\end{figure}

\noindent
The TF diagram of the local calibrators is shown in
Fig.\ref{fig:TFcalib_PT_r}.
The dispersion is $0.42$\,mag. The slope is consistent with that of
Pierce \& Tully (1992), although we found inconsistency in $B$
(\S\ref{sec:THE TULLY-FISHER RELATION Calibration}). This is because the
$r$-band TF diagram in Pierce \& Tully (1992) shows
no appreciable difference in slope between
$\log W_{20}>2.30$ and $\log W_{20}<2.30$, which is not the case in the
$B$-band. Willick et al. (1997) gave
$7.73$ for the slope, which also matchs well with ours. Our correction
for $A_i$ and for $A_G$ follows the same schemes as theirs, thus the
zero point of the TF calibration is also comparable with ours. Their
zero point is $-20.12$ ($H_0=65$\,\kmsMpc\ assumed), which is in good
agreement with ours.

\begin{figure}
\vspace*{-15mm}
\begin{center}
\epsfig{file=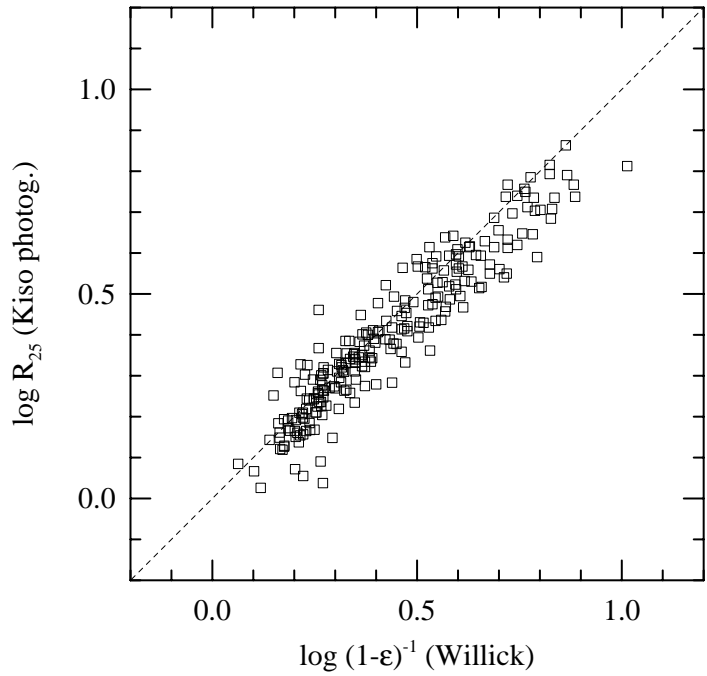,
        width=350pt, height=300pt, 
	bbllx=70pt, bburx=770pt, bblly=20pt, bbury=620pt, 
	clip=, angle=0}
\end{center}\vspace{-5mm}
\caption{A comparison of the major-to-minor axial ratios between the
	 data of Willick (1991) and ours.
	 \label{fig:logRWil_vs_logR25}}
\end{figure}

A systematic difference of $\Delta\log R_{25}=0.03$ is found between
our axial ratio $R_{25}$ and that measured by Willick (1991)
(Fig.\ref{fig:logRWil_vs_logR25}).
This difference introduces errors of 0.03-0.45\,mag in $A_i$ and
0.3-7\,\% in $W_{20}$ depending on galaxy morphology and inclination,
respectively. A comparative calculation using both sets of the axial
ratios for galaxies common between the $r$- and the $B$-band samples
shows that our axial ratios gives $H_0$ which is 3\% smaller than
that in $r$.

\begin{figure}
\begin{center}
\epsfig{file=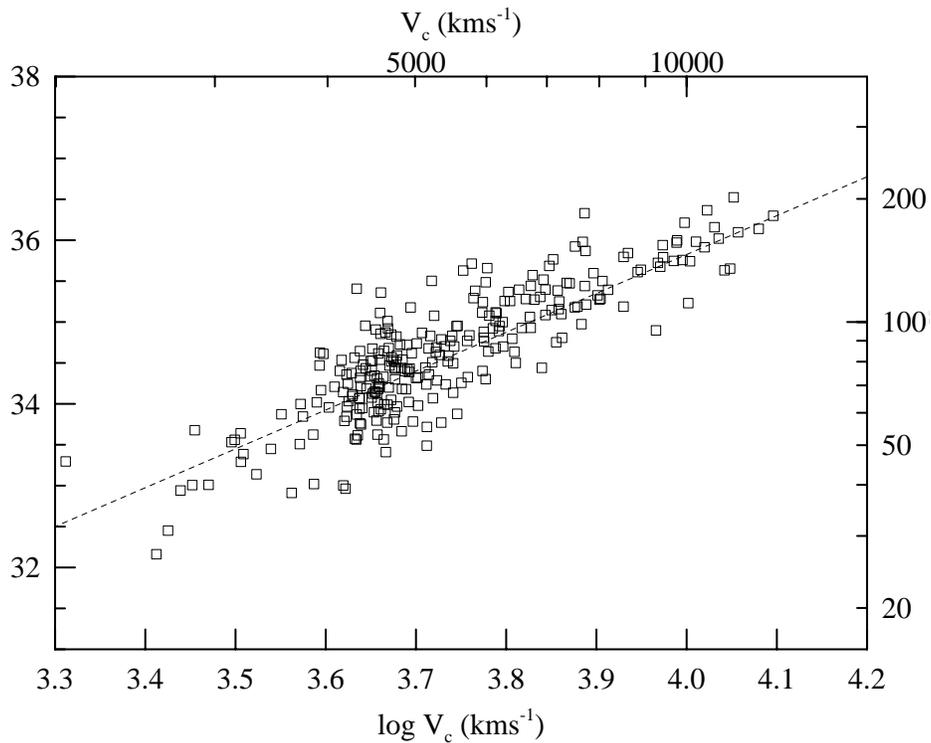,
        width=350pt, height=300pt, 
	bbllx=70pt, bburx=770pt, bblly=20pt, bbury=620pt, 
	clip=, angle=0}
\end{center}\vspace{-5mm}
\caption{The velocity-TF distance relation for the $r$-band TF sample.
	 A dashed line represents an arbitrary
	 Hubble's law with $\log h=-0.15$.
	 \label{fig:logV_vs_DMKisoWil}}
\end{figure}

The velocity-distance relation for the $r$-band TF sample is shown in
Fig.\ref{fig:logV_vs_DMKisoWil}.
We find that the $r$-band sample shows tighter velocity-distance
correlation (correlation coefficient = 0.873) than for
$\mu_{T\!F}^{(B)}$
(correlation coefficient = 0.804, Fig.\ref{fig:logV_vs_DM_449}). This
may be accounted for by the following two reasons. First, the $r$-band
TF relation shows smaller intrinsic scatter than $B$-band (Pierce and
Tully 1988). Second, the CCD observations for the $r$-band sample
provide more accurate photometry than our photographic photometry.

\begin{figure}
\vspace*{-15mm}
\begin{center}
\epsfig{file=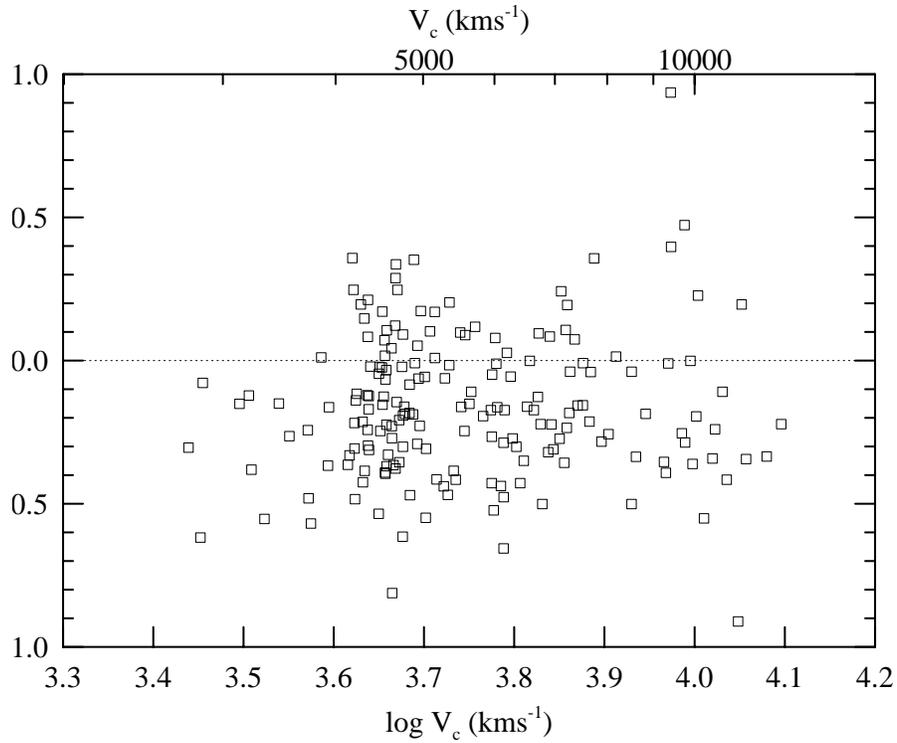,
        width=350pt, height=300pt, 
	bbllx=70pt, bburx=770pt, bblly=20pt, bbury=620pt, 
	clip=, angle=0}
\end{center}\vspace{-5mm}
\caption{Difference between the TF distance moduli $\mu_{T\!F}^{(B)}$
         and $\mu_{T\!F}^{(r)}$.
	 \label{fig:logV_vs_dDM_Br_Binc}}
\end{figure}

Difference between $\mu_{T\!F}^{(r)}$ and $\mu_{T\!F}^{(B)}$ is plotted
against $\log V_c$ in Fig.\ref{fig:logV_vs_dDM_Br_Binc}.
There is a systematic difference of 0.16\,mag. If this is ascribed to
the Malmquist bias, i.e.,
$1.38(\sigma_{T\!F}^{(B)\;2}-\sigma_{T\!F}^{(r)\;2})=0.16$\,mag, then we
obtain $\sigma_{T\!F}^{(r)}=0.35$\,mag from
$\sigma_{T\!F}^{(B)}=0.49$\,mag (\S\ref{sec:RESULT Global Value}),
This value is indeed in
agreement with what we obtain in the following maximum likelihood
analysis (\S\ref{sec:ANALYSIS WITH WILLICK'S r-BAND SAMPLE Result}).

Note that the $r$-band calibrators miss NGC\,925 and 4536 which show
relatively large deviation in $B$ from the calibrated relation (Table
\ref{tbl:B-band local calibrators}). It is possible that future
inclusion of these galaxies considerably changes our $r$-band result.

\subsection{Result}
\label{sec:ANALYSIS WITH WILLICK'S r-BAND SAMPLE Result}

\begin{figure}
\begin{center}
\epsfig{file=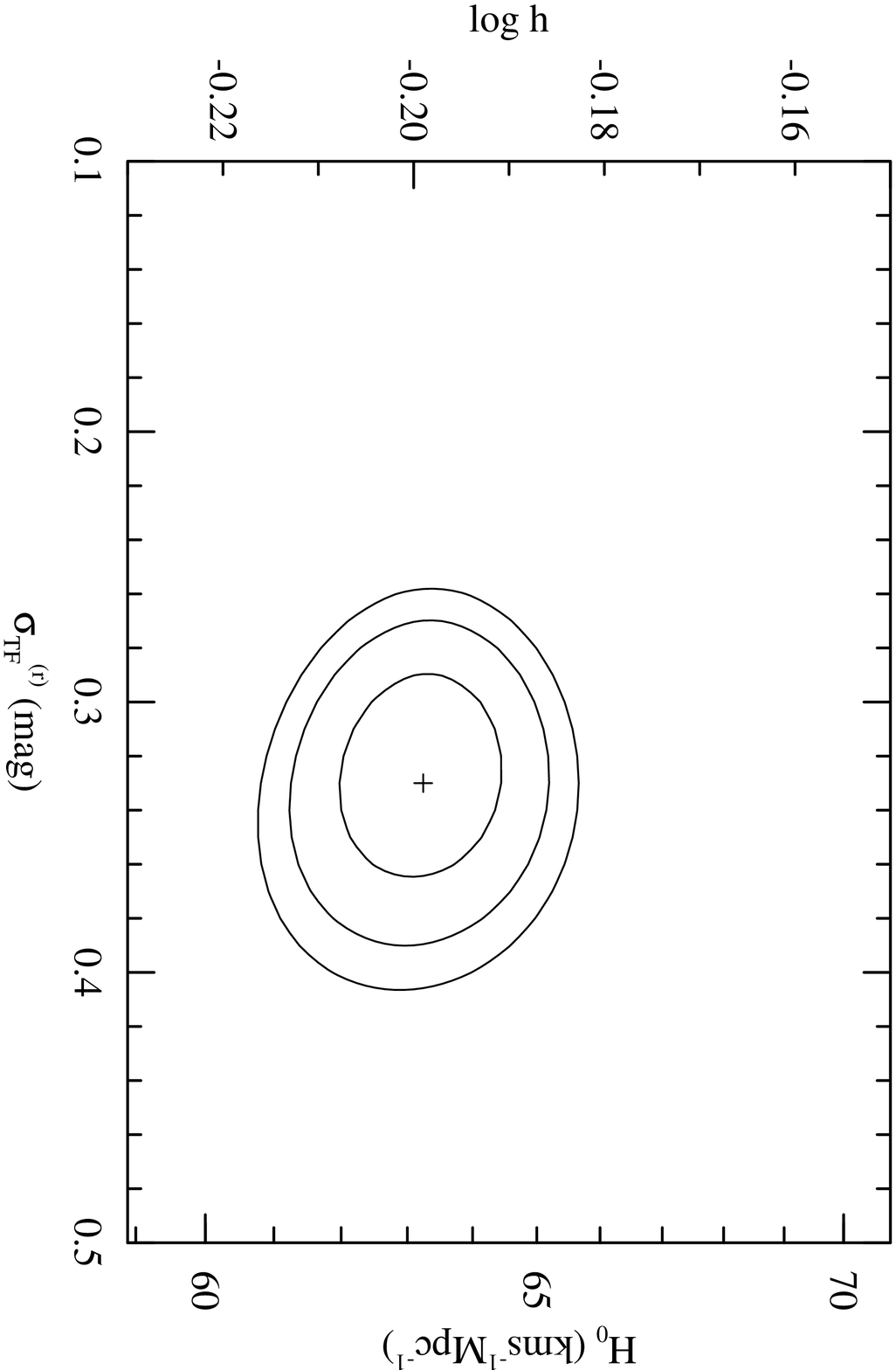,
        width=350pt, height=300pt, 
	bbllx=70pt, bburx=770pt, bblly=20pt, bbury=620pt, 
	clip=, angle=0}
\end{center}\vspace{-5mm}
\caption{A probability contour map for $H_0$ and $\sigma_{T\!F}^{(r)}$
	 obtained from the $r$-band TF sample.
	 Confidence levels at 70\%, 95\%
	 and 99\% are shown. \label{fig:sTF_vs_logh_rcomB218Binc}}
\end{figure}

A contour map for $H_0$ and $\sigma_{T\!F}^{(r)}$ obtained from the
maximum likelihood analysis is shown in
Fig.\ref{fig:sTF_vs_logh_rcomB218Binc}. The internal and external errors
are assessed in the same way as for the $B$-band sample. We finally
obtain

\begin{equation}
H_0=63\pm 1^{\;+13}_{\;-\;\,7}\,\mbox{\kmsMpc}\,.
\label{eqn:H0_rband}
\end{equation}

\noindent
The quoted errors have the same meaning as equation (\ref{eqn:H0_400}).
The intrinsic scatter of the $r$-band TF relation is
obtained to be $\sigma_{T\!F}^{(r)}=0.33\pm 0.05$\,mag. This
$\sigma_{T\!F}^{(r)}$ well explains the systematic difference between
$\mu_{T\!F}^{(r)}$ and $\mu_{T\!F}^{(B)}$ shown in
Fig.\ref{fig:logV_vs_dDM_Br_Binc}.

\begin{figure}
\begin{center}
\epsfig{file=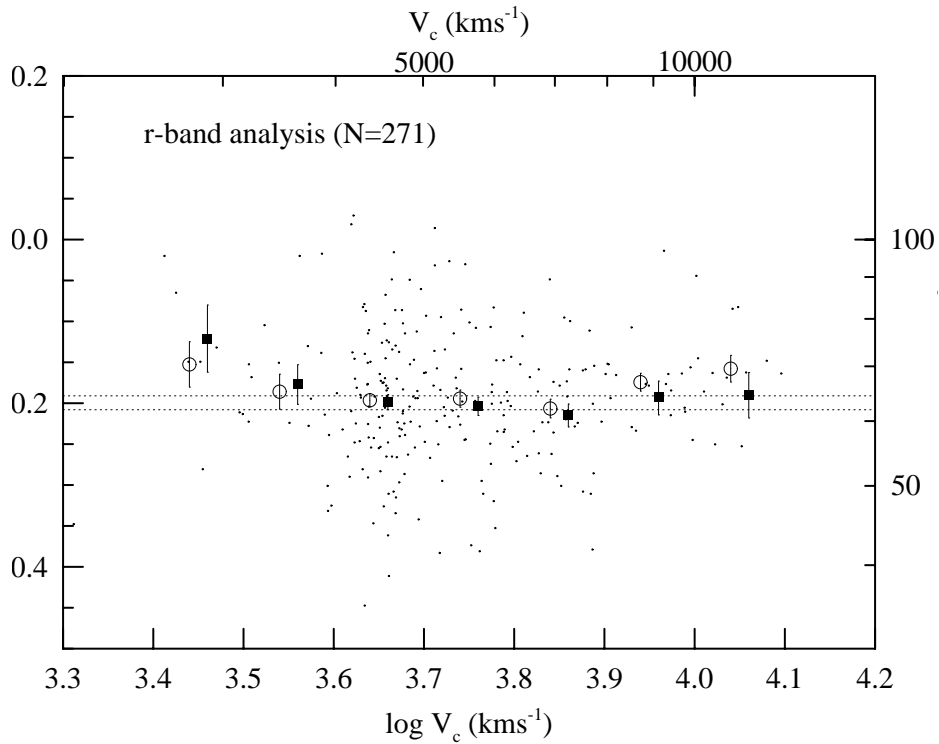,
        width=350pt, height=300pt, 
	bbllx=70pt, bburx=770pt, bblly=20pt, bbury=620pt, 
	clip=, angle=0}
\end{center}\vspace{-5mm}
\caption{Spatial variation of $H_0$ obtained from the $r$-band TF
         sample.
	 The intrinsic scatter $\sigma_{T\!F}^{(r)}$ is fixed for all
	 the subsamples in $\log V_c$ bins to 0.33\,mag obtained in
	 Fig.19. Dots represent the individual galaxies, to which
	 the Hubble ratio $V_c/r_{T\!F}^{(r)}$ is given as $H_0$. Filled
	 squares represent the most probable $H_0$ in each bin obtained
	 by the maximum likelihood method, while open circles represent
	 the sample average of $\log(V_c/r_{T\!F}^{(r)})$. The squares
	 and
	 the circles are shifted from the center of each $\log V_c$ bin
	 to avoid the overlap. Error bars are taken from the 70\%
	 error obtained for each subsample. Two dashed lines indicate
	 the 70\% confidence levels $H_0=62$ and 64\,\kmsMpc\ shown
	 in Fig.19.
          \label{fig:logV_vs_logh_rcomB218Binc}}
\end{figure}

The spatial variation of $H_0$ is shown in
Fig.\ref{fig:logV_vs_logh_rcomB218Binc}. Except for poor subsamples at
$\log V_c<3.5$, the values of $H_0$'s are mostly in agreement with the
global value within their errors. Apparent
increase of a sample mean of $\log(V_c/r_{T\!F}^{(r)})$ toward larger
$\log V_c$ is smaller than that for $B$-band analysis
(Fig.\ref{fig:logV_vs_logh_B441}). This is consistent with
$\sigma_{T\!F}^{(r)}$ which is smaller than $\sigma_{T\!F}^{(B)}$.

\section{DISCUSSION} \label{sec:DISCUSSIONS}
\subsection{Global Value of $H_0$} \label{sec:DISCUSSIONS Global Value}

We compare our result with those obtained by other studies from the TF
relation, the period-luminosity relation of Cepheid variables,
surface brightness fluctuation (SBF), the planetary nebulae luminosity
function (PNLF), and supernovae (SNe) type Ia and \II . Extensive
reviews for these methods are found in Jacoby et al. (1992), van den
Bergh (1994), Livio, Donahue and Panagia (1997), and Okamura (1998).

\subsubsection{The \TF\ relation}\label{sec:DISCUSSIONS TF relation}

IF92 obtained
$H_0=78.5_{\;-\;\,9.1}^{\;+10.3}$\,\kmsMpc\ using the TF relation
calibrated with five calibrators (NGC\,224, 598, 300, 2403, 3031). If
we adopt the same calibrators as theirs but new Cepheid distances, we
obtain a TF calibration of
$M^{(B)}_{T\!F}=-19.44-6.19\,(\log W_{20}-2.5)$. The slope matchs that
of IF92, $-6.24$, but the zero point is 0.28\,mag brighter. Using this
calibration, we obtain
$H_0=74\pm 2^{\;+17}_{\;-10}$\,\kmsMpc .
The difference of $H_0$ between IF92 and this calculation is fully
ascribed to the TF zero point difference of 0.28\,mag.

\begin{figure}
\vspace*{-30mm}
\begin{center}
\epsfig{file=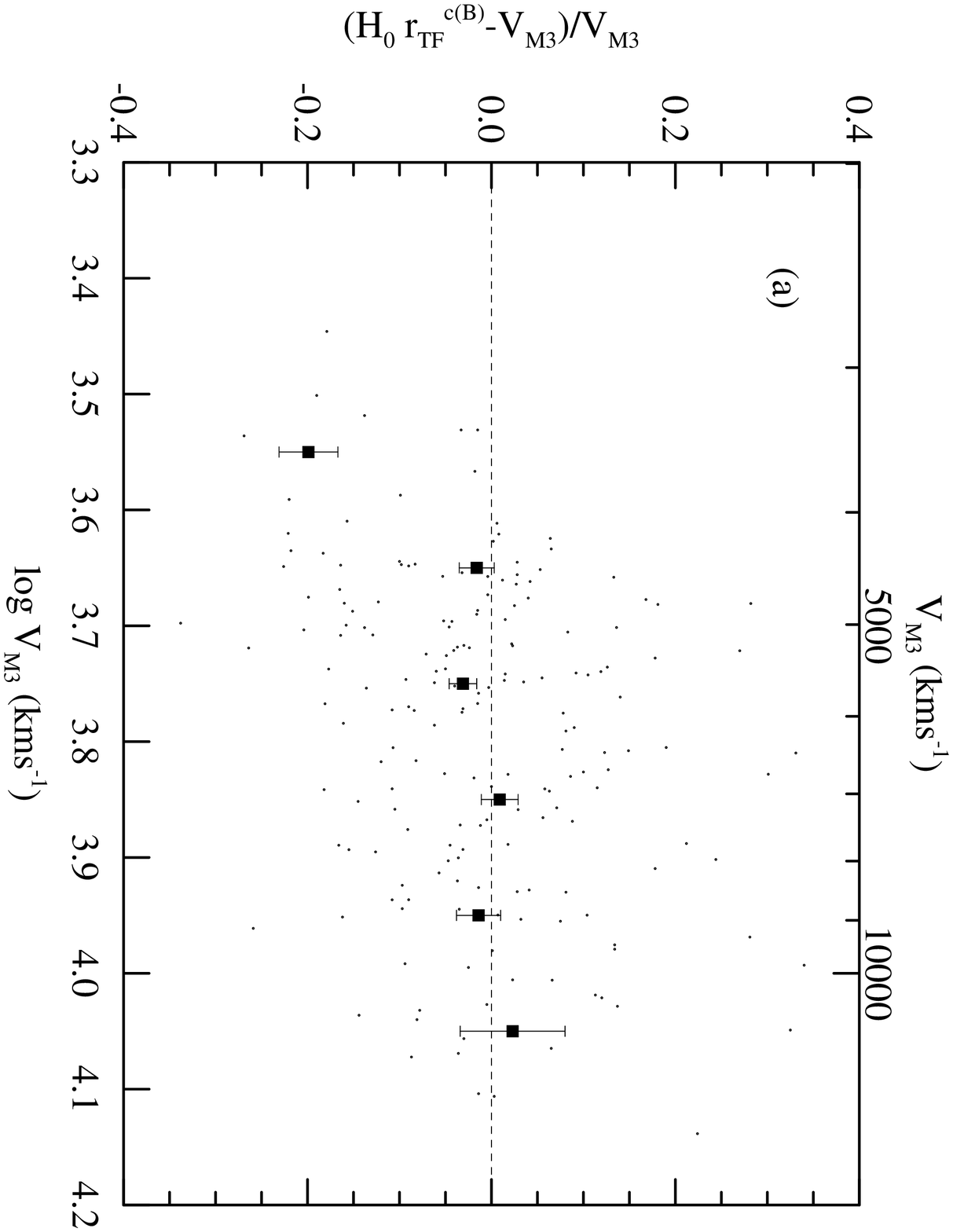,
        width=350pt, height=300pt, rheight=280pt,
	bbllx=70pt, bburx=770pt, bblly=20pt, bbury=620pt, 
	clip=, angle=0}
\epsfig{file=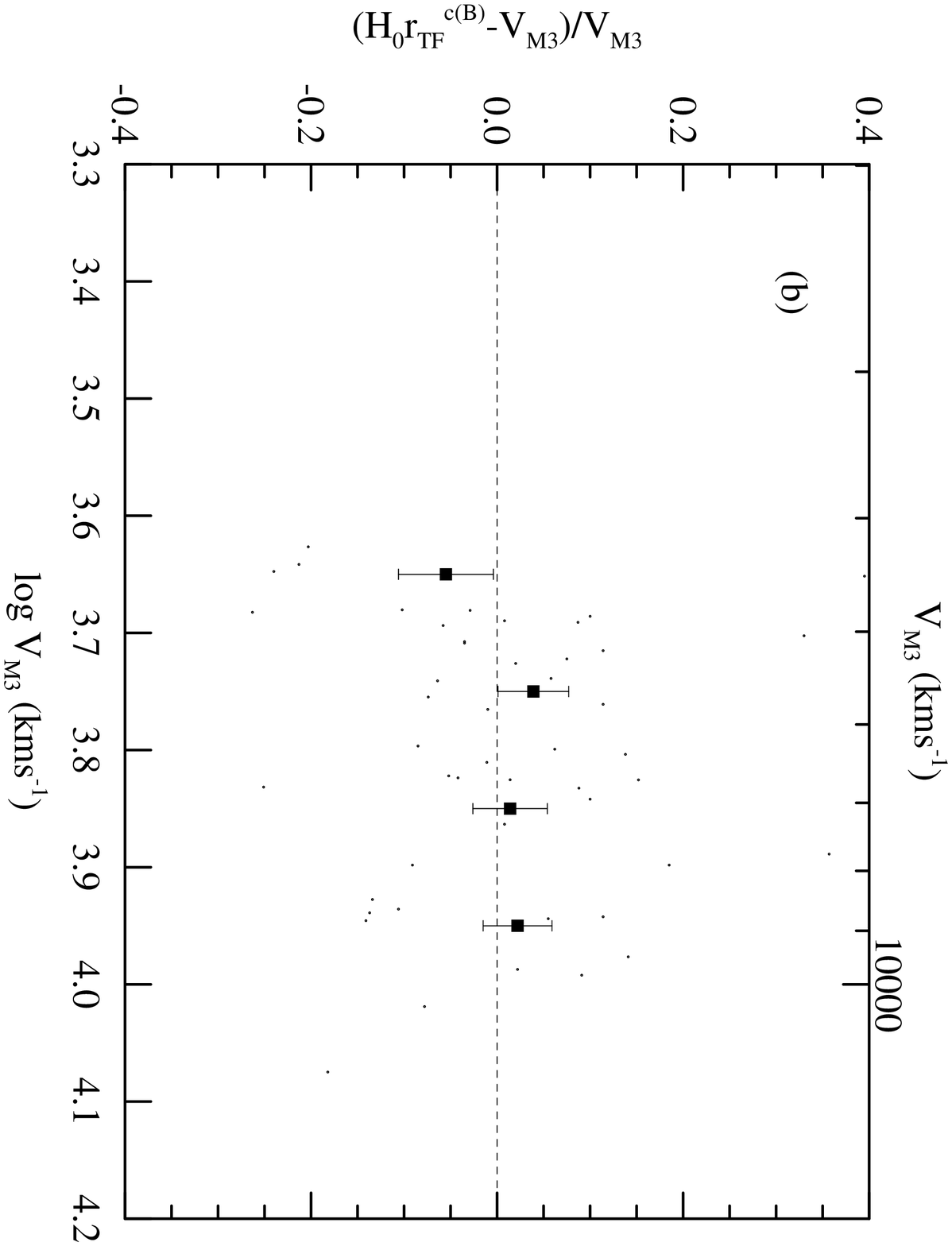,
        width=350pt, height=300pt, 
	bbllx=70pt, bburx=770pt, bblly=20pt, bbury=620pt, 
	clip=, angle=0}
\end{center}\vspace{-5mm}
\caption{Relative difference of the Hubble velocity between Mark
	 \III\ catalog (Willick et al. 1997) and ours, plotted against
	 the Mark \III\ Hubble velocity. (a) ``w91pp''  (b) ``hmcl''.
	 Dots represent individual galaxies and solid squares show
	 median in each $\log V_{M\!3}$ bin. Error bars are calculated
	 as a standard deviation divided by a square root of the number
	 of galaxies in each bin.
	 \label{fig:Vmark3_vs_Vdiffratio}}
\end{figure}

The Mark \III\ catalog (Willick et al. 1997) provides Hubble velocities
of spirals using the TF relation. For our galaxies common to the Mark
\III\ ``w91pp'' and ``hmcl'' samples, we compute the Hubble
velocities using $H_0^{(B)}=65$\,\kmsMpc\ and our TF distance moduli
$r_{T\!F}^{c(B)}$ corrected for the Malmquist bias. The result of
comparison is shown in
Fig.\ref{fig:Vmark3_vs_Vdiffratio}. No systematic difference is found in
the Hubble velocity, except for a bin $3.5\le \log V_{M\!3}<3.6$ of
w91pp.
For this bin, we observe $H_0$ which is larger than those of other
bins (Fig.\ref{fig:logV_vs_logh_B441}). This locally larger value of
$H_0$ is compatible with the difference of the Hubble velocity in this
bin. These facts imply an actual infall motion of galaxies in the bin
toward the \PP\ supercluster.

\begin{figure}
\vspace*{-15mm}
\begin{center}
\epsfig{file=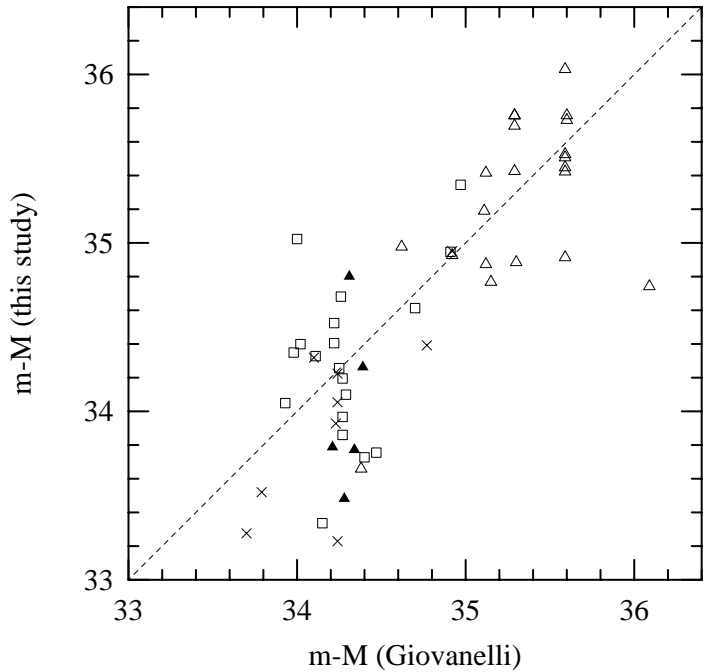,
        width=350pt, height=300pt, 
	bbllx=70pt, bburx=770pt, bblly=20pt, bbury=620pt, 
	clip=, angle=0}
\end{center}\vspace{-5mm}
\caption{Comparison of the distance modulus between Giovanelli et al.
	 (1997b) and this study. Squares, crosses, closed and open
	 triangles represent galaxies in NGC\,383 group, NGC\,507 group,
	 A\,262, and A\,2634/A\,2666, respectively. Distance moduli for
	 their galaxies are based on $H_0=69$\, \kmsMpc\ from Giovanelli
	 et al. (1997a). \label{fig:DMG97_vs_DM}}
\end{figure}

Recently, Giovanelli et al. (1997a) studied 24 clusters of galaxies out
to $V_c\sim 9000$\,\kms\ and obtained $H_0=69\pm 5$\,\kmsMpc . Our
result $H_0^{(B)}=65$\,\kmsMpc\ is in good agreement with theirs. Four
of their clusters reside in the \PP\ region we observed. A comparison
of the galaxy distances is shown in Fig.\ref{fig:DMG97_vs_DM}. Their
distance modulus here is converted from their original recession
velocity (Giovanelli et al. 1997b) based on $H_0=69$\,\kmsMpc . The
systematic difference is $-0.06$\,mag, indicating a satisfactory
agreement in the distances. A vertical spread is apparent at
$m-M=34.4$.
The spread, $\Delta (m-M)\sim 0.5$\,mag, is consistent with our TF
dispersion $\sigma_{T\!F}^{(B)}=0.49\pm 0.05$\,mag.

\subsubsection{Other Methods} \label{sec:DISCUSSIONS Other methods}

The period-luminosity relation of Cepheid variables, SBF, and PNLF are
currently available to galaxies up to only a few tens of Mpc. Using
Cepheids in galaxies in the Virgo cluster, Pierce et al. (1994) and
Freedman et al. (1994) obtained $H_0=87\pm 7$ and
$80\pm 17$\,\kmsMpc , respectively. The error quoted by Pierce
et al. (1994) did not include the cluster depth effect for which
Freedman et al. (1994) assumed $\pm 0.35$\,mag. If the same depth effect
is taken into account in the result of Pierce et al. (1994), the total
error becomes $\pm 17$\,\kmsMpc . Tanvir et al. (1995)
identified Cepheid variables in a galaxy in the Leo I group and obtained
a smaller value $H_0=69\pm 8$\,\kmsMpc . More recently, Freedman (1998)
presented $H_0=73\pm 14$\,\kmsMpc\ in an interim report of the HST key
project.
From the SBF method with a new zero point calibrated with Cepheid
observations, Tonry et al. (1997) obtained
$H_0=81\pm 6\,$\kmsMpc . Using the PNLF method,
McMillan, Ciardullo and Jacoby (1993) obtained from three galaxies in
the Fornax cluster $H_0=81\pm 8$\,\kmsMpc\ and
$69\pm 8$\,\kmsMpc , which depend on the choice of the Fornax/Coma
distance ratio of 5.25 (Faber et al. 1989) and 6.14 (Aaronson et al.
1989), respectively. Although PNLF distances are based on the M31
distance of 710\,kpc instead of 770\,kpc which we adopt, the difference
reduces their $H_0$'s by only 3\%.

Using a Hubble diagram of 13 SNe type Ia calibrated with a single
Cepheid distance to SN\,1972E in NGC\,5253, Riess, Press and Kirshner
(1995) obtained $H_0=67\pm 7$\,\kmsMpc\ within $V_c\sim
30000$\,\kms . Hamuy et al. (1996) obtained
$H_0=63.1\pm 3.4\pm 2.9$\,\kmsMpc\ in a similar spatial extent using a
Hubble diagram constructed with 29 SNe of type Ia and calibrated with
four SNe (SN\,1972E in NGC\,5253, SN\,1937C in IC\,4182, and SN\,1981B
in NGC\,4536, and SN\,1990N in NGC\,4639) to which Cepheid distances
are available. These $H_0$'s are both in good agreement with ours
(see also Saha et al. 1995, Tammann and Sandage 1995, Saha et al. 1996).
The expanding photosphere method has been developed for the distance
determination with type \II\ SNe (Kirshner and Kwan 1974). The method
directly provides the distance to the SNe without any external
calibration procedures. Schmidt et al. (1994) obtained
$H_0=73\pm 6\pm 7\,$\kmsMpc\ using sixteen SNe type \II\ extending
to $V_c\sim 15000$\,\kms . Their result is slightly larger than but in
marginal agreement with ours.

Shanks (1997) compared TF distances to 11 spirals with their Cepheid
distances and to 12 spirals with their SNe Ia distances. The TF
distances were taken from Pierce (1994). They concluded that, overall,
the TF distance is smaller by $0.43\pm 0.11$\,mag than the Cepheid/SNe
distances. This leads to that $H_0$ obtained by Pierce (1991) using
the Virgo and the Ursa Major clusters
is revised downwards from $H_0=84\pm 10$\,\kmsMpc\ to
$H_0=69\pm 8$\,\kmsMpc . This $H_0$ is in good agreement with ours.

From the above overview, our $H_0=65\pm 2^{\;+20}_{\;-14}$\,\kmsMpc\
agrees with the results by Cepheids observations, the SNe Ia and \II\,
and the Tully-Fisher relation. Although sample galaxies are sparse,
Cepheids and SNe observations currently provide us with the most
reliable distances in the local Universe ($\sim 20\,h^{-1}$\,Mpc) and
to much further galaxies ($\sim 500\,h^{-1}$\,Mpc), respectively. It is
a good demonstration that our estimate of the global $H_0$ is consistent
with both of these results obtained from such different fields of the
Universe. It is also convincing that our $H_0$ is in agreement with the
result of an independent TF study, Giovanelli et al. (1997a), which used
sparse but widely distributed galaxies. Further comparative studies of
the Cepheid, SBF, and PNLF distances will work out the current
inconsistency of $H_0$ among these three methods. Then, Cepheid
observations of distant galaxies will give a definite conclusion for the
global value of $H_0$.

\subsection{Bulk Motion} \label{sec:DISCUSSIONS peculiar motions}

A bulk motion in the Pisces-Perseus region was investigated by Willick
(1990), Han and Mould (1992), Courteau et al. (1993), Baffa et
al. (1993), and Hudson et al. (1997). Willick (1990) claimed that the
\PP\ supercluster and the galaxies behind the supercluster up to
$V_c\sim 8000\,$\kms\ showed a negative coherent flow. The result
was supported by Courteau et al.(1993), who argued that all galaxies
within a sphere out to at least $V_c\sim 6000\,$\kms\ in radius
around the Local Group showed parallel streaming flow with amplitude of
$\sim -400\,$\kms\ toward the Great Attractor. Han and Mould (1992)
studied clusters of galaxies lying beyond the \PP\ supercluster, and
also detected a negative coherent motion of the clusters.
Baffa et al. (1993) reported that galaxies around the \PP\
supercluster are infalling into the supercluster, i.e., that the
supercluster is shrinking. Most recently, Hudson et al.(1997) studied
seven clusters in the Pisces-Perseus region and found peculiar motions
which are consistent with no bulk motion in the CMB frame. They also
argued on the basis of $\chi^2$-statistics that their result is also
consistent with the negative bulk motion found by Courteau et al.
(1993).

\begin{figure}
\begin{center}\vspace{-20mm}
\epsfig{file=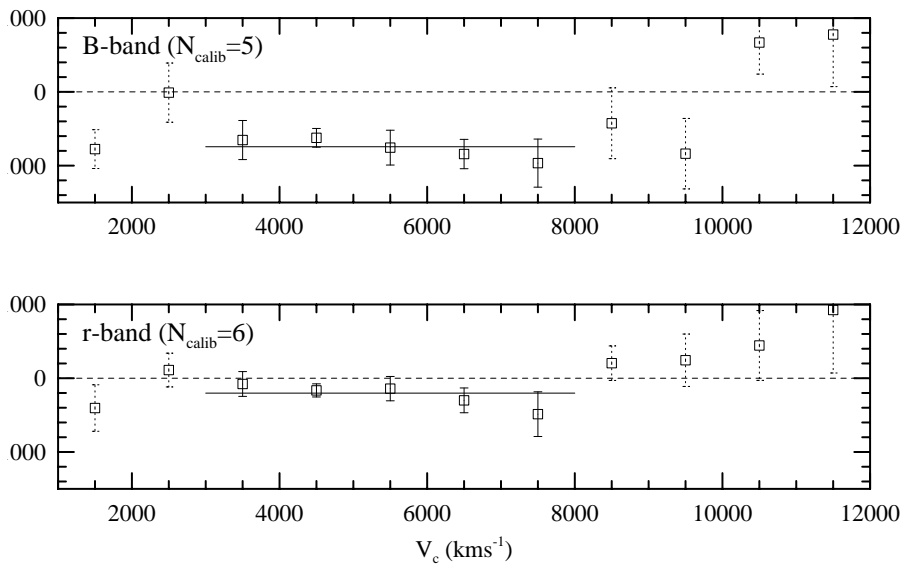,
        width=350pt, height=300pt, 
	bbllx=70pt, bburx=770pt, bblly=20pt, bbury=620pt, 
	clip=, angle=0}
\end{center}\vspace{-15mm}
\caption{Peculiar velocities plotted against the CMB-rest recession
	 velocity in $B$ (upper) and $r$ (lower). A solid horizontal
	 line shows a mean value ($V_p^{(B)}=744$\,\kms\ in $B$ and
	 $V_p^{(r)}=202$\,\kms\ in $r$) in the range
	 $3000\le V_c\le 8000$\,\kms . \label{fig:Vc_vs_Vpec}}
\end{figure}

We show in Fig.\ref{fig:Vc_vs_Vpec} mean peculiar velocities derived
from our $H_0$ and $r_{T\!F}^c$ corrected for the Malmquist bias.
Because of the small number of galaxies at $V_c<3000$\,\kms , we discard
these galaxies from the current discussion. At a farther region,
$V_c>8000$\,\kms , we find a serious effect of the sampling
incompleteness
bias (see Fig.\ref{fig:logV_vs_logh_B441}). The same effect is found in
Courteau et al. (1993) and Mathewson and Ford (1994), but they managed
to circumvent the effect by comparing the peculiar velocity with that of
a control field. As we have no such control data, we simply restrict
ourselves to the range of $3000\le V_c\le 8000$\,\kms\ for the current
discussion on the peculiar motion.

In the $B$-band ($H_0^{(B)}=65$\,\kmsMpc ) we find a mean negative bulk
flow of $V_p^{(B)}=-744\pm 92$\,\kms , while a flow with a smaller
amplitude $V_p^{(r)}=-202\pm 68$\,\kms\ is obtained in $r$
($H_0^{(r)}=63$\,\kmsMpc ). If we change $V_c$ of each galaxies in
$3000\le V_c\le 8000$\,\kms\ to
$V_c-V_p^{(B)}$ and then re-calculate $H_0$, we obtain
$H_0=72$\,\kmsMpc . This value is significantly larger than
$H_0^{(B)}=65$\,\kmsMpc . This naive correction of $V_c$ for
$V_p^{(B)}$ is probably invalid, however, since it substantially reduces
the likelihood. The ratio of the likelihood for $H_0^{(B)}=65$\,\kmsMpc\
to that for $H_0=72$\,\kmsMpc\ is $2.6\times 10^{-6}$, which indicates
less
than 0.001\% probability for the corrected $V_c$ being well fitted with
the Hubble flow. A 95\% probability interval of $V_p^{(B)}$
inferred from the likelihood ratio is
$-290\le V_p^{(B)}\le +470$\,\kms , each of the boundary
corresponding to $H_0=68$ and $61$\,\kmsMpc , respectively.
The inconsistency between the 95\% probability interval and
$V_p^{(B)}=-744$\,\kms\
may be
caused by the large uncertainty of the $B$-band TF distances due to a
large $\sigma_{T\!F}^{(B)}$. It also implies that the current $B$-band
TF analysis is not suitable for deducing accurate peculiar motions for
individual galaxies.

In $r$-band, on the other hand,
correction of $V_c$ for $V_p^{(r)}=-202$\,\kms\ leads to
$H_0=65$\,\kmsMpc\ which is not significantly different from
$H_0^{(r)}=63$\,\kmsMpc . Further, a 95\% probability interval of
$V_p^{(r)}$
inferred from the likelihood ratio is
$-510\le V_p\le +140$\,\kms , which is consistent with the mean
negative bulk flow $V_p^{(r)}=-202$\,\kms .
Boundaries of the 95\% probability interval of $V_p^{(r)}$ corresponds
to $H_0=68$ and $62$\,\kmsMpc , respectively.
This consistent result
is convincing for the existence of the negative bulk flow.
The range of $V_p$ is consistent with those of the earlier results
(Table.\ref{tbl:Comparison}). We consider that the $r$-band TF
analysis for the peculiar motions seems more reliable than that in
$B$-band.
Our result $H_0\sim 65$\,\kmsMpc\ is also consistent with negative
bulk flows which were obtained in the previous studies {\em
independently from an absolute value of $H_0$}\/.

\section{SUMMARY} \label{sec:SUMMARY}

We perform an unbiased estimate of the global $H_0$ using a TF relation
with a maximum likelihood method. We use a homogeneous,
magnitude-limited sample of 441 galaxies ($m_z\le 15.5$\,mag)
distributed up to $cz\sim 12000$\,\kms\ in the \PP\ region. The
reliability of the method was examined by the Monte-Carlo simulation,
and it was shown that the method works
satisfactorily in deriving an unbiased $H_0$ for the present sample. The
results of the present analysis are summarized as follows.
\begin{enumerate}
\item We obtain
$H_0^{(B)}=65\pm 2^{\;+20}_{\;-14}$\,\kmsMpc\ for the unbiased estimate
of the global value of the Hubble constant.
The first and second error
terms represent the internal and external errors, respectively. All the
known sources of systematic errors are taken into account in the
estimate of the external error. The same
analysis with the $r$-band TF sample of 271 galaxies gives
$H_0^{(r)}=63\pm 1_{\;-\;\,7}^{\;+13}$\,\kmsMpc , which is
consistent with $H_0^{(B)}$. These $H_0$'s are in good agreement with
IF92 if we allow for the difference of the TF calibration. We also find
a good agreement for these $H_0$'s with those obtained via Cepheid
observations for local Universe (Freedman 1998), the Tully-Fisher
relation (Giovanelli et al. 1997a) and supernovae (Riess,
Press and Kirshner 1995, Hamuy et al. 1996) applied to a spatial volume
larger than $\sim 8000$\,\kms .
\item A bulk motion in the \PP\ region is briefly investigated using
our $H_0$.
Analysis of the $r$-band data compiled by Willick et al. (1997)
gives a negative motion of
$V_p\sim -200$\,\kms\ between the range $3000\le V_c\le 8000$\,\kms . The amplitude of
$V_p$ is consistent with the 95\% probability interval
$-510\le V_p\le +140$\,\kms\ inferred from the likelihood ratio.
The interval is consistent with the bulk flows obtained by Willick
(1990), Han and Mould (1992), Courteau et al. (1993), and Hudson et al.
(1997). Analysis of the $B$-band data does not give a meaningful result
because of a larger intrinsic dispersion of the TF relation.
\end{enumerate}

\acknowledgments

We are grateful to the past and the present staff of the Kiso
observatory for taking a number of photographic plates of a good
quality. We thank M. Doi, K. Shimasaku, N. Yasuda, N. Kashikawa, and M.
Fukugita for their stimulating discussions and valuable suggestions at
an early stage of this work. A part of the literature data was taken
from Astronomical Data Analysis Center (ADAC) at National Astronomical
Observatory of Japan (NAOJ). A part of this work was financially
supported in 1994 and 1995 fiscal years by the Research Followships of
the Japan Society for the Promotion of Science for Young Scientists. We
also thank an anonymous referee for many useful suggestions and comments
which largely improved an earlier version of this paper.

\clearpage

\begin{table}[h]
\begin{center}
\begin{tabular}{llr} \tableline \tableline
& & \multicolumn{1}{c}{\raisebox{-.3ex}[0mm]{Number of}} \\
\multicolumn{2}{c}{\raisebox{1.7ex}[0mm]{Criteria}} & \multicolumn{1}{c}{\raisebox{.5ex}[0mm]{galaxies}} \\
\tableline
\multicolumn{2}{l}{Photometry sample}   & 1524\hspace{1.5eM} \\
& & \\
\multicolumn{2}{l}{21cm data available} &  957\hspace{1.5eM} \\
& & \\
Morphological type & Sa-Sd              &  754\hspace{1.5eM} \\
Inclination & $i\ge 45^\circ$        &  523\hspace{1.5eM} \\
Recession velocity & $V_c\ge 1000\,$\kms   &  514\hspace{1.5eM} \\
21cm line-profile width & $2.30\le \log W_{20}<2.75$   &  $^{(a)}$449\hspace{1.5eM} \\
Absolute magnitude residual & $|M_{T\!F}^{(B)}-M_B|<1.62\,$mag &  $^{(b)}$441\hspace{1.5eM} \\
\tableline
\end{tabular}
\end{center}
\caption{Sample selection for the Tully-Fisher analysis
	 \label{tbl:Sample selection for the TF analysis}}
\tablecomments{(a) Spiral sample. (b) TF sample.}
\end{table}

\begin{table}[h]
\begin{center}
\begin{tabular}{ l@{\ }  l
		 l@{} l@{}  l@{} l@{}l@{}l@{}l
		 r@{.}l@{ (.}l@{)\hspace{1eM}}
		 r@{.}l@{ (.}l@{)\hspace{1eM}}
		 c
		 c
		 r
		 r@{.}l
		 r@{.}l
		 r@{.}l } \tableline \tableline
\multicolumn{2}{@{}c@{}}{\raisebox{-1.ex}[0mm]{Name}} &
\multicolumn{7}{@{}c@{}}{\raisebox{-1.0ex}[0mm]{Type}} &
\multicolumn{3}{@{}c@{}}{\raisebox{-.6ex}[0mm]{$\mu$}} &
\multicolumn{3}{@{}c@{}}{\raisebox{-.6ex}[0mm]{$B_T^0$}} &
\multicolumn{1}{@{}c@{}}{\raisebox{-.6ex}[0mm]{$r_T^0$}} &
\multicolumn{1}{@{}c@{}}{\raisebox{-.6ex}[0mm]{$W_{20}^{obs}$}} &
\multicolumn{1}{@{}c@{}}{\raisebox{-.6ex}[0mm]{$V_h$}} &
\multicolumn{2}{@{}c@{}}{\raisebox{-1.ex}[0mm]{$\log R_{25}$}} &
\multicolumn{4}{   c@{}}{\raisebox{-.6ex}[0mm]{$M_B-M_{T\!F}^{(B)}$}} \\
\multicolumn{2}{@{}c@{}}{\raisebox{ .2ex}[0mm]{    }} &
\multicolumn{7}{@{}c@{}}{\raisebox{ .2ex}[0mm]{    }} &
\multicolumn{3}{@{}c@{}}{\raisebox{ .2ex}[0mm]{{\small [mag]}}} &
\multicolumn{3}{@{}c@{}}{\raisebox{ .2ex}[0mm]{{\small [mag]}}} &
\multicolumn{1}{@{}c@{}}{\raisebox{ .2ex}[0mm]{{\small [mag]}}} &
\multicolumn{1}{@{}c@{}}{\raisebox{ .2ex}[0mm]{{\small [\kms ]}}} &
\multicolumn{1}{@{}c@{}}{\raisebox{ .2ex}[0mm]{{\small [\kms ]}}} &
\multicolumn{2}{@{}c@{}}{\raisebox{ .2ex}[0mm]{             }} &
\multicolumn{4}{   c@{}}{\raisebox{ .2ex}[0mm]{{\small [mag]}}} \\
\multicolumn{2}{@{}c@{}}{\raisebox{.5ex}[0mm]{\small (1)}} &
\multicolumn{7}{@{}c@{}}{\raisebox{.5ex}[0mm]{\small (2)}} &
\multicolumn{3}{@{}c@{}}{\raisebox{.5ex}[0mm]{\small (3)}} &
\multicolumn{3}{@{}c@{}}{\raisebox{.5ex}[0mm]{\small (4)}} &
\multicolumn{1}{@{}c@{}}{\raisebox{.5ex}[0mm]{\small (5)}} &
\multicolumn{1}{@{}c@{}}{\raisebox{.5ex}[0mm]{\small (6)}} &
\multicolumn{1}{@{}c@{}}{\raisebox{.5ex}[0mm]{\small (7)}} &
\multicolumn{2}{@{}c@{}}{\raisebox{.5ex}[0mm]{\small (8)}} &
\multicolumn{2}{@{}c@{}}{\raisebox{.5ex}[0mm]{\small (9)}} &
\multicolumn{2}{@{}c@{}}{\raisebox{.5ex}[0mm]{\small (10)}} \\
\tableline
N224$^\dagger$ & (M31) &
.    & S   & A     & S & 3 & . & . &
$^{(a)}$24   & 44  & 13 &
3    & 34  & 02 & 2.67 &
536  &
$-300$\ \ &
\ 0  & 49 &
0 & 13 & $-0$ & 01 \\
N598$^\dagger$ &  (M33) &
. & S & A & S & 6 & . & . &
$^{(b)}$24 & 64 & 09 &
5 & 74 & 03 & 5.49 &
199 &
$-179$\ \ &
0 & 23 &
0 & 11 & $0$ & 09 \\
N300  &        &
. & S & A & S & 7 & . &. &
$^{(c)}$26 & 66 & 10 &
8 & 53 & 05 & 7.89 &
150 &
142\ \ &
0 & 15 &
0 & 48 & 0 & 49 \\
N2403 &        &
. & S & X & S & 6 & . & . &
$^{(d)}$27 & 51 & 24 &
8 & 41 & 07 & 7.91 &
244 &
131\ \ &
0 & 25 &
0 & 39 & 0 & 34 \\
N3031$^\dagger$& (M81) &
.    & S   & A  & S  &2  &.  & . &
$^{(e)}$27 & 80 & 20 &
7 & 40 & 03 & 6.36 &
434 &
$-34$\ \ &
0 & 28 &
0 & 55 & 0 & 42 \\
N925  &       &
.    & S   & X  & S  &7  &.  & . &
$^{(f)}$29 & 84 & 16 &
10 & 00 & 11 & --- &
215 &
$553$\ \ &
0 & 25 &
$-0$ & 69 & $-0$ & 72 \\
N3368 & (M96) &
.    & S   & X  & T  &2  &.  & . &
$^{(g)}$30 & 32 & 16 &
9 & 86 & 13 & --- &
354 &
$897$\ \ &
0 & 16 &
0 & 38 & 0 & 26 \\
N4536 &       &
.    & S   & X  & T  &4  &.  & . &
$^{(h)}$31 & 10 & 13 &
10 & 56 & 08 & --- &
336 &
$1804$\ \ &
0 & 37 &
$-0$ & 45 & $-0$ & 53  \\
N1365$^\dagger$&       &
.    & S   & B  & S  &3  &.  & . &
$^{(i)}$31 & 32 & 17 &
 9 & 90 & 07 & 8.94 &
398 &
$1662$\ \ &
0 & 26 &
$-0$ & 67 & $-0$ & 79    \\
N4639$^\dagger$&       &
.    & S   & X  & T  &4  &.  & . &
$^{(j)}$32 & 00 & 23 &
11 & 92 & 10 & --- &
356 &
$1010$\ \ &
0 & 17 &
0 & 72 & 0 & 60  \\
\tableline
\end{tabular}
\end{center}
\caption{Local calibrators \label{tbl:B-band local calibrators}}
\tablecomments{Col(2): Morphological type taken from RC3.
Col(3): Distance modulus and its error.
$\mu=18.50\,$mag is assumed for LMC.
References are:
($a$) Freedman and Madore (1990),
($b$) Freedman, Wilson and Madore (1991),
($c$) Freedman et al. (1992),
($d$) Freedman and Madore (1988),
($e$) Freedman et al. (1994),
($f$) Silbermann et al. (1996),
($g$) Tanvir et al. (1995),
($h$) Saha et al. (1996),
($i$) Freedman, Madore and Kennicutt (1997), and
($j$) Sandage et al. (1997).
Col(4): $B$-band magnitude corrected for the Galactic and the
internal extinction and its error taken from
RC3. Col(5): $r$-band magnitude corrected for the Galactic and the
internal extinction prescribed by Willick et al. (1997).
Col(6): Observed 21cm line-profile width taken
from RC3. Col(7): Heliocentric recession velocity
taken from RC3. Col(8): Logarithmic major-to-minor axial ratio taken
from RC3. Col(9),(10): Difference between the $B$-band absolute
magnitude and that derived from the Tully-Fisher relation calibrated
with five (col.9) and ten (col.10) calibrators. $^\dagger$ See text.}
\end{table}

\clearpage

\begin{table}[h]
\begin{center}
\begin{tabular}{lr@{\,}cr@{\,}ccr@{\,}cr@{\,}cr@{\,}c} \tableline\tableline
& \multicolumn{4}{@{}c@{}}{$A_G$} & & \multicolumn{6}{@{}c@{}}{$A_i$} \\
\cline{2-5}\cline{7-12}
& \multicolumn{2}{@{}c@{}}{RC2$-$BH} & \multicolumn{2}{@{}c@{}}{RSA$-$BH} & &
  \multicolumn{2}{@{}c@{}}{RC2$-$RC3} & \multicolumn{2}{@{}c@{}}{RSA$-$RC3} &
  \multicolumn{2}{@{}c@{}}{Bothun$-$RC3} \\
\tableline
TF sample & 0.14 & (.06) & $-0.07$ & (.07) & &
	      $-0.12$ & (.07) & 0.36 & (.15) & $-0.17$ & (.06) \\
Local calibrators & 0.15 & (.05) & $-0.06$ & (.03) & &
	      $-0.14$ & (.07) & 0.31 & (.14) & $-0.15$ & (.03) \\
Difference & $-0.01$ & & $-0.01$ & & & 0.02 & & 0.05 & & $-0.02$ & \\
\tableline
\multicolumn{12}{l}{\footnotesize{Dispersion is given in parentheses.}}
\end{tabular}
\end{center}
\caption{Difference of average extinction corrections for the TF sample
	 \label{tbl:correction difference}}
\end{table}

\begin{table}[h]
\begin{center}
\begin{tabular}{cl@{\,}lcc} \tableline \tableline
\\
\multicolumn{5}{@{}c@{}}{\raisebox{1.ex}[0mm]{\underline{\hspace{1ex}$\circ$ Internal random error\hspace{1ex}}}}  \\
& $\cdot$ Statistics (Fig.\ref{fig:sTF_vs_logh_B441}) & ($\pm2$\,\kmsMpc )           & $\pm2.7$\% &  \\
\\
\multicolumn{5}{@{}c@{}}{\raisebox{1.ex}[0mm]{\underline{\hspace{1ex}$\circ$ External systematic error\hspace{1ex}}}} \\
& $\cdot$ \TF\ calibration & ($\pm0.20$\,mag)        & $\pm9.7$\% \\
& $\cdot$ LMC distance & ($\pm0.1$\,mag)             & $\pm4.6$\% \\
& $\cdot$ Photometric zero point & ($\pm0.05$\,mag)  & $\pm2.3$\% \\
& $\cdot$ Axial ratio & ($\Delta\log R_{25}=0.03$) & $+9$\% \\
& $\cdot$ Galactic extinction & ($\pm0.01$\,mag)     & $\pm0.5$\% \\
& $\cdot$ Internal extinction & ($\pm0.05$\,mag)     & $\pm2.3$\% \\
& $\cdot$ Parameters ($\sigma_{rand}$, $\Delta W_{20}^{obs}$, $\Delta\log R_{25}$, $\Delta B_T$) & ($\Delta\log h=0.005$)    & $\pm1.2$\% \\
\cline{2-4}
& \multicolumn{2}{@{}c@{}}{(Arithmetic sum)} & $+30$\%, $-21$\% \\
\\
\tableline
\end{tabular}
\end{center}
\caption{Summary of the errors for $H_0$ \label{tbl:error summary}}
\end{table}

\begin{table}[h]
\begin{center}
\begin{tabular}{llr@{}c@{}rr@{}c@{}rr@{}c@{}r} \tableline \tableline
&&&&& \multicolumn{3}{c}{\raisebox{-.3ex}[0mm]{Volume}} &
\multicolumn{3}{c}{\raisebox{-.3ex}[0mm]{$V_p$}} \\
\multicolumn{2}{c}{\raisebox{2.5ex}[0mm]{Reference}} & \multicolumn{3}{c}{\raisebox{2.5ex}[0mm]{R.A.}} &
\multicolumn{3}{c}{\raisebox{1ex}[0mm]{(\kms )}} &
\multicolumn{3}{c}{\raisebox{1ex}[0mm]{(\kms )}} \\
\tableline
This work & {\small ($B$, $H_0=68-61$\,\kmsMpc )} & $22^h$ & -- & $4^h$ & 3000 & -- & 8000 & \multicolumn{3}{c}{$(-290, +470)$} \\
------------ & {\small ($r$, $H_0=63$\,\kmsMpc )} & $22^h$ & -- & $4^h$ & 3000 & -- & 8000 & \multicolumn{3}{c}{$-202\pm 68$} \\
------------ & {\small ($r$, $H_0=68-62$\,\kmsMpc )} & $22^h$ & -- & $4^h$ & 3000 & -- & 8000 & \multicolumn{3}{c}{$(-510, +140)$} \\
\multicolumn{2}{l}{Willick (1990)}                      & $22^h$ & -- & $4^h$ & 3800 & -- & 6000 & \multicolumn{3}{c}{$-441\pm 49$} \\
\multicolumn{2}{l}{Han and Mould (1992)}                & $0^h$ & -- & $2^h$ & 4000 & -- & 7500 & \multicolumn{3}{c}{$-420\pm 115$} \\
\multicolumn{2}{l}{Courteau et al. (1993)}              & $22^h$ & -- & $4^h$ & 4000 & -- & 9000 & \multicolumn{3}{c}{$-383\pm 58$} \\
\multicolumn{2}{l}{Hudson et al. (1997)}                & $0^h$ & -- & $4^h$ & 4000 & -- & 6000 & \multicolumn{3}{c}{$-60\pm 220$} \\
\tableline
\end{tabular}
\end{center}
\caption{Comparison with previous works of the bulk flow in the
	 Pisces-Perseus region \label{tbl:Comparison}}
\end{table}

\clearpage

\renewcommand{\arraystretch}{1.2}

\clearpage

\clearpage

\clearpage

\clearpage

\clearpage

\clearpage

\clearpage

\clearpage

\clearpage

\clearpage

\clearpage

\clearpage

\clearpage

\vspace{-15mm}

\clearpage

\clearpage

\clearpage

\end{document}